\documentclass[final,5p,times,twocolumn,nonatbib]{elsarticle}

\usepackage{amssymb}
\usepackage{amsmath}
\usepackage{eurosym}
\usepackage{multirow}

\usepackage{booktabs}
\usepackage{graphicx}

\usepackage{subcaption}
\usepackage{enumitem}

\PassOptionsToPackage{hyphens}{url}

\usepackage[hidelinks]{hyperref}
\usepackage{orcidlink}
\usepackage[final]{microtype}
\usepackage[
backend=biber,
style=numeric,
sorting=none,
date=year,
doi=true,
url=true,
eprint=false,
isbn=false
]{biblatex}

\AtEveryBibitem{%
	\clearfield{note}%
	\clearfield{urldate}%
	\clearfield{urlyear}%
	\clearfield{urlmonth}%
	\clearfield{urlday}%
}
\journal{Computer Networks}

\begin{document}

\begin{frontmatter}



\title{Usability of Passwordless Authentication in Wi-Fi Networks: A Comparative Study of Passkeys and Passwords in Captive Portals}

\author[rnasa,citic]{Martiño Rivera-Dourado~\orcidlink{0000-0003-4301-9417}}

\author[rnasa,citic]{Rubén Pérez-Jove~\orcidlink{0000-0002-7988-945X}}

\author[rnasa,citic]{Alejandro Pazos~\orcidlink{0000-0003-2324-238X}}

\author[rnasa,citic]{Jose Vázquez-Naya~\orcidlink{0000-0002-6194-5329}}

\affiliation[rnasa]{organization={RNASA-IMEDIR, Universidade da Coruña},
	addressline={Facultade de Informática, Campus de Elviña}, 
	city={A Coruña},
	postcode={15071}, 
	state={Galicia},
	country={Spain}}

\affiliation[citic]{organization={Centro de Investigación CITIC, Universidade da Coruña},
	addressline={Campus de Elviña}, 
	city={A Coruña},
	postcode={15071}, 
	state={Galicia},
	country={Spain}}


\begin{abstract}
Passkeys have recently emerged as a passwordless authentication mechanism, yet their usability in captive portals remains unexplored. This paper presents an empirical, comparative usability study of passkeys and passwords in a Wi-Fi hotspot using a captive portal. We conducted a controlled laboratory experiment with 50 participants following a split-plot design across Android and Windows platforms, using a router implementing the FIDO2CAP protocol. Our results show a tendency for passkeys to be perceived as more usable than passwords during login, although differences are not statistically significant. Independent of the authentication method, captive portal limitations negatively affected user experience and increased error rates. We further found that passkeys are generally easy to configure on both platforms, but platform-specific issues introduce notable usability challenges. Based on quantitative and qualitative findings, we derive design recommendations to improve captive portal authentication, including the introduction of usernameless authentication flows, improved captive portal detection mechanisms, and user interface design changes.

\end{abstract}


\begin{keyword}


passkeys \sep captive portals \sep usability study \sep FIDO2 \sep WebAuthn \sep network authentication

\end{keyword}

\end{frontmatter}

\flushbottom



\section{Introduction}
\label{sec:intro}
Public Wi-Fi hotspots are widely used to provide internet connectivity in dense environments, such as airports, stadiums, hospitals, retail establishments, and public transportation. According to the latest Cisco Annual Internet Report, in 2023 there were an estimated 628 million public Wi-Fi hotspots worldwide~\cite{cisco_cisco_2020}.

To control connectivity in public Wi-Fi hotspots, captive portals~\cite{larose_captive_2020} are often deployed to block access to network resources. The fundamental mechanism of a captive portal involves intercepting initial traffic and redirecting the user to a web server hosting the portal. Beyond controlling simple internet access, captive portals enable additional functionality, such as enforcing acceptable use policies, facilitating payment, conducting surveys, or supporting user authentication~\cite{marques_eap-sh_2020}.

When applied to Wi-Fi access control, captive portals typically employ web-based authentication mechanisms, such as OAuth-based social logins or traditional username/password schemes. However, password-based authentication is vulnerable to various security threats, including phishing, credential guessing, and data breaches. Although prior work has aimed to improve password usability~\cite{tan_practical_2020}, passwords are often reused across multiple services or forgotten by users, introducing significant security and usability challenges.


To address the limitations of passwords~\cite{bonneau_quest_2012}, the FIDO Alliance has introduced passkeys as a replacement for web authentication. These public-key credentials rely on an authenticator device, which can be an external security key (e.g. a YubiKey) or a platform authenticator integrated into the user device (e.g. Windows Hello).

Passkeys mitigate many common attacks against passwords, including phishing, server-side credential compromise, and keylogging. The underlying technology is described in the FIDO2 standards: the W3C WebAuthn API and the FIDO Client-to-Authenticator Protocol (CTAP). WebAuthn defines the interface and protocol interactions between the web server and the authenticator via the web browser. At a lower level, CTAP defines how clients and platforms communicate with FIDO2 authenticators, where passkeys are stored. Together, these standards ensure interoperability and compatibility among FIDO2 components.


Although passkeys are increasingly adopted in web applications, they have not yet been widely implemented in network access control systems~\cite{rivera-dourado_eap-fido_2025}. In this context, the FIDO2 Captive Portal Authentication Protocol (FIDO2CAP)~\cite{rivera-dourado_novel_2024} was designed to bring passkeys to network authentication. FIDO2CAP adapts the captive portal reference architecture~\cite{larose_captive_2020} to support FIDO2 authentication using a web browser, a FIDO2 authentication server, and a FIDO2 authenticator. With this approach, users can access networks using a previously registered device, such as a laptop or smartphone.

Although FIDO2 authentication is already supported by current end-user devices, several usability challenges arise when adopting FIDO2 in captive portals. First, the captive portal mini-browsers embedded in operating systems~\cite{wang_capturing_2023} often lack compatibility with the WebAuthn API, forcing implementations to direct users to a compatible browser. Second, the FIDO2 authentication flow requires users to perform additional, unfamiliar steps, including interacting with the authenticator (user presence) and performing local authentication (user verification). Both issues negatively affect the user experience in Wi-Fi hotspots.


The usability of FIDO2 authentication has been studied in various contexts, such as online accounts~\cite{reynolds_tale_2018}, enterprise environments~\cite{kepkowski_challenges_2023}, and general web applications~\cite{lyastani_is_2020}. Unlike general web applications, captive portals require users to authenticate before gaining network connectivity, which typically involves interaction with operating system network settings and embedded browser contexts. These additional steps make authentication more tedious to repeat and add cognitive load for users who connect frequently. Although FIDO2 authentication has been studied in web applications, the user experience of FIDO2-based authentication in captive portals remains underexplored, given the recent emergence of passkeys and their application to network access systems. Building on this prior work, we hypothesize that passkeys achieve better usability metrics than passwords in captive portals, due to reduced cognitive load and the removal of password management overhead.

To fill this research gap, this paper presents an empirical, comparative usability study of passkeys and passwords as authentication methods in a Wi-Fi hotspot using a captive portal. Specifically, we define the following research questions:

\begin{itemize}
	\item \textbf{RQ1:} Are passkeys easier to use compared to passwords during login in a captive portal?
	\item \textbf{RQ2:} Is it easy for users to configure passkeys on their devices?
	\item \textbf{RQ3:} How can a captive portal improve its usability for authentication?
\end{itemize}

As mentioned, our hypothesis for RQ1 is that passkeys achieve better usability metrics than passwords during the login process in captive portals. For RQ2, our hypothesis is that users find it easy to configure passkeys on their devices, given the increasing support for FIDO2 in modern operating systems. After the experiment, for RQ3, we identified specific usability challenges in captive portal authentication and derived recommendations to enhance the user experience.

\vspace{5px}

Accordingly, the main contributions of this paper are: 
\begin{itemize}
	\item An implementation of FIDO2CAP adapted to support both passwords and passkeys, integrated with the Zitadel Identity and Access Management (IAM) system for easy configuration and management.
	\item A laboratory experiment with $N=50$ participants examining user experience with passkeys and passwords in captive portals, using a split-plot design across Android and Windows platforms. This experiment resulted in quantitative and qualitative findings of the task effectiveness, efficiency, and the user satisfaction.
	\item Design recommendations to improve captive portal authentication based on the observed usability challenges and user feedback.
\end{itemize}

Our results suggest that passkeys tend to be perceived as more usable than passwords during login, although differences are not statistically significant. Independent of the authentication method, captive portal limitations negatively affected user experience and increased error rates. We found that passkeys are generally easy to configure on both platforms. However platform-specific issues, such as Android captive portal mini-browser incompatibility with the WebAuthn API, introduce notable usability challenges. Based on quantitative and qualitative findings, we derive design recommendations to improve captive portal authentication, including the introduction of usernameless authentication flows, improved captive portal detection mechanisms, and user interface design changes.

\section{Background on FIDO2}
\label{sec:background}

FIDO2~\cite{angelogianni_how_2024} is an open authentication standard developed by the Fast Identity Online (FIDO) Alliance and the World Wide Web Consortium (W3C). It is composed of two specifications: (1) the WebAuthn protocol, which defines the registration and authentication ceremonies and provides a client interface for Relying Parties (RPs) to interact with user authenticators; and (2) the Client-to-Authenticator Protocol (CTAP), which enables communication between the client and the authenticator.

Registration and authentication in WebAuthn are based on the public-key cryptographic design of FIDO2 credentials. During the registration ceremony, the authenticator is requested to generate a new credential (i.e. a passkey), composed of a public key and a private key. The private key is stored securely inside the authenticator, while the public key is sent to the server. Later, during the authentication ceremony, the server sends a challenge to the authenticator, which signs it using the private key. The server verifies the signature, completing the authentication process.

Private keys for passkeys are stored within authenticators, a key component of the FIDO2 standard. These devices keep credentials secure and provide a protected environment for cryptographic operations. These FIDO2-conformant authenticators can be external and roaming~\cite{grossklags_security_2017} (e.g. a security key via USB, Bluetooth, or NFC) or internal, protected by a TPM or a Trusted Execution Environment (TEE), as in platform authenticators on many smartphones.

To prevent unauthorized use, authenticators can implement two levels of protection: User Presence (UP) and User Verification (UV). UP is a basic protection measure, primarily against malware: the user must consent to the registration or authentication operation, usually by pressing a physical button. Optionally, Relying Parties (RPs) can require UV, which obligates the user to locally authenticate to consent to the operation. UV can be achieved via a PIN or biometrics, such as facial or fingerprint recognition.

Some FIDO2 client and browser implementations support cross-device authentication. This feature allows a user to authenticate on one device using a passkey registered on another device. A common scenario is using a passkey registered on a smartphone to authenticate to a website on a laptop. To achieve cross-device authentication, the browser on the laptop displays a QR code that the smartphone scans, enabling a local connection between devices (typically via Bluetooth). After pairing, the user initiates authentication on the laptop and consents to the operation (UP/UV) on the smartphone, which acts as the FIDO2 authenticator.

Original FIDO protocols were conceived primarily for two-factor authentication: users first authenticate with a knowledge factor (e.g., a password) and then use a FIDO authenticator as the second factor. Later, FIDO2 passkeys introduced the support for passwordless authentication, where a passkey can serve as the sole factor. In addition, FIDO2 enabled a even more simplified flow for usernameless authentication, in which users do not have to enter a username before signing in. Instead, the RP can request credential discovery, allowing the compatible FIDO2 authenticator to select an eligible passkey and thereby identify the user during the authentication ceremony.

\section{System overview}
\label{sec:system-overview}

In \cite{rivera-dourado_novel_2024} we introduced the FIDO2 Captive Portal Authentication Protocol (FIDO2CAP), designed to bring passkeys to network authentication by using captive portals. This section provides an overview of the system used in the usability experiment, based on FIDO2CAP, with several additional functionalities to improve user experience and administration capabilities.

Captive portals are familiar to users. Once they connect to the network via the operating system settings, the FIDO2CAP system opens the captive portal in a compatible web browser. As in any captive portal, the connectivity to network resources is disabled until the user is correctly authenticated. For users to authenticate to the network, the user portal displays the login page of  the  WebAuthn Authentication Web Application (WAWA).

FIDO2CAP is composed of the user equipment, the FIDO2 authenticator, the captive portal modules, user and session databases. The authentication occurs by challenging the FIDO2 authenticator, which may be a roaming hardware security key or a platform authenticator, via a web browser within the user equipment. This communication relies on the user equipment compatibility with the WebAuthn API and the FIDO CTAP2 protocols. 

In this work, we compare the usability of passwords and passkeys as authentication methods in captive portals. Therefore, the system must implement both authentication flows, and should be easy to configure. To achieve this and to provide a familiar User Experience (UX), the FIDO2CAP system was integrated with the Zitadel Identity and Access Management (IAM), a feature rich free and open source software. 

\begin{figure}[ht!]
	\centering
	\includegraphics[width=\linewidth]{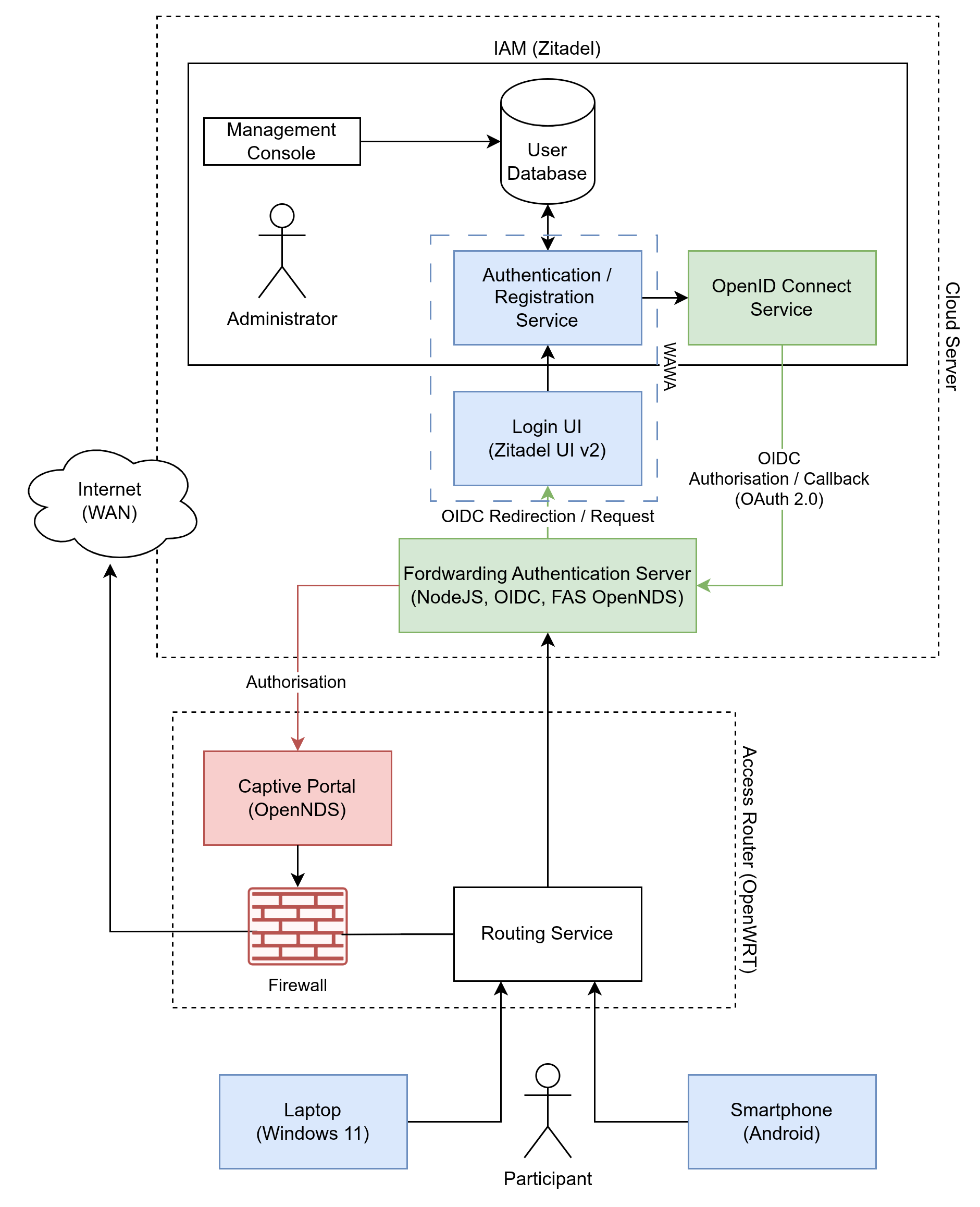}
	\caption{Architecture of the captive portal infrastructure used in the laboratory experiment.}
	\label{fig:system-overview:architecture}
\end{figure}

The final system architecture is shown in Figure~\ref{fig:system-overview:architecture}. Specifically, we integrated the OpenNDS-compatible Forwarding Authentication Server (FAS) with Zitadel via OpenID Connect (OIDC), powered by OAuth 2.0. Thanks to this integration, an administrator can easily configure access policies and manage users. Additionally, the system used in this work incorporates a customised version of the Login UI v2 of Zitadel, which provides the possibility to implement two independent registration and authentication flows: one based on passwords, and the other on passkeys via WebAuthn (FIDO2). This UX feature allowed us to compare both authentication methods in a between-groups usability experiment.

The environment where the experiments were conducted is composed of end-devices and the captive portal infrastructure installed in a Wi-Fi network. Below, we describe the software configured and installed on each device in the experimental environment:

\begin{itemize}
	\item \textbf{User end-devices.} Wi-Fi clients with Captive Portal Detection (CPD) and a browser compatible with the WebAuthn API to support FIDO2 authentication. Both devices feature FIDO2 platform authenticators, allowing users to store passkeys within the operating system, protected by the device lock.
	\begin{enumerate}
		\item Laptop, powered by Windows 11.
		\item Smartphone, powered by Android 14.
	\end{enumerate}
	\item \textbf{Captive Portal infrastructure.} 
	\begin{enumerate}
		\item Wi-Fi access router, powered by OpenWRT, configured to work with FIDO2CAP using the OpenNDS captive portal.
		\item Cloud server, powered by Ubuntu and Docker, with the following deployed microservices.
		\begin{itemize}
			\item HTTP reverse proxy (NGINX), configured to manage requests to the different microservices.
			\item IAM (Zitadel), exposing the API and the web application for managing policies and users.
			\item Login UI (Zitadel UI v2), serving the front end of the login page powered by the Zitadel IAM server and acting as the front end of the WAWA server.
			\item Forwarding Authentication Server (FAS), managing requests from the OpenNDS captive portal and connected to the Zitadel IAM via OIDC.
		\end{itemize}
	\end{enumerate}
\end{itemize}

\begin{figure}[ht!]
	\centering
	\begin{subfigure}[b]{0.30\linewidth}
		\centering
		\includegraphics[width=\linewidth]{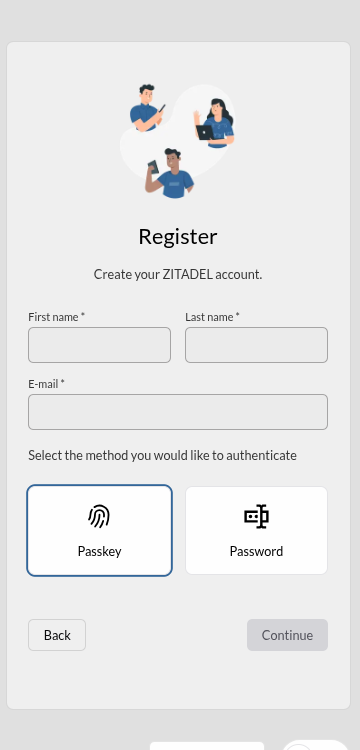}
		\caption{Registration form.}
	\end{subfigure}
	\hfill
	\begin{subfigure}[b]{0.30\linewidth}
		\centering
		\includegraphics[width=\linewidth]{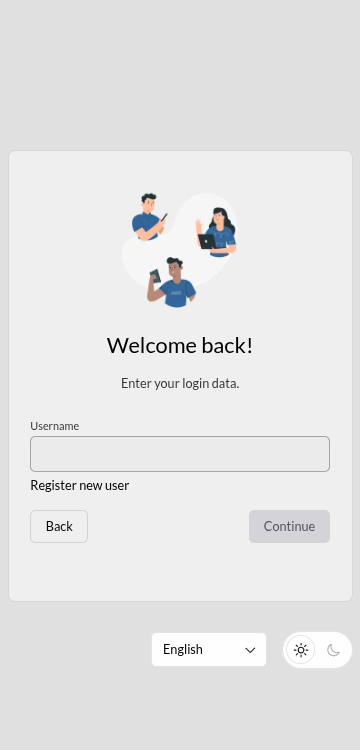}
		\caption{Username input.}
	\end{subfigure}
	\hfill
	\begin{subfigure}[b]{0.30\linewidth}
		\centering
		\includegraphics[width=\linewidth]{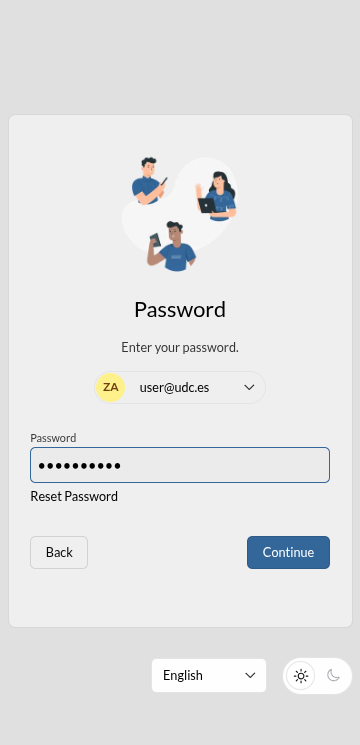}
		\caption{Password input.}
	\end{subfigure}
	\caption{Login user interface in Zitadel, used in the captive portal.}
	\label{fig:zitadel-ui-screenshots}
\end{figure}

The operation of this infrastructure is based on the FIDO2CAP protocol, integrated with the Zitadel IAM and the login interface. When the user connects to the Wi-Fi router, OpenNDS provides the user portal of the WAWA server. This page then redirects the user to the login UI using OIDC. Powered by the Zitadel IAM, the login UI authenticates the user with their registered method (password or passkey) and replies to the WAWA server via OIDC. Finally, the WAWA server authorises the user to access the network, and the OpenNDS enforcement device (a firewall) enables connectivity for the user device. Notably, there is a variable delay between authorisation by the WAWA server and the OpenNDS firewall rule update, due to network latency and processing time inherent to the captive portal system. Some screenshots of the WAWA user interface are included in Figure~\ref{fig:zitadel-ui-screenshots}.

\section{Methodology}
\label{sec:methods}

The goal of the experiment is to study the usability of passkey authentication in captive portals and compare it with password authentication in the same scenario. In addition, the study aims to assess the usability of passkey registration and use on two different devices: a Windows laptop and an Android smartphone.

This section describes the experiment design, including the procedure and the data collection and analysis.

\subsection{Experiment design}
\label{subsec:methods:experiment-design}


The experiment is designed as a split-plot study in a controlled laboratory, where the captive portal system and the end devices are deployed (see Section~\ref{sec:system-overview}).

Participants were divided into two groups: $G1_{\text{passkey}}$ and $G2_{\text{password}}$. Members of $G1_{\text{passkey}}$ were instructed to register and authenticate with passkeys, while members of $G2_{\text{password}}$ used passwords. This design allows us to study the usability of the authentication factor as a between-group variable.

Within each group, participants performed the tasks on two end devices: the Windows laptop and the Android smartphone. This within-subject factor captures usability of configuration and use across devices. Participants may experience learning effects when performing tasks on the second device; therefore, the order of device exposure was counterbalanced (see Section~\ref{subsec:methods:experiment-procedure}).

Participants were instructed to perform two unassisted tasks on each end device: registration and authentication. First, to initiate registration, the participant clicks on the shortcut to the registration portal, which guides them to create a user and configure the assigned method. Second, the participant connects to the Wi-Fi network via the captive portal using the credentials they have just configured. The next subsection details both tasks and the stages of the experiment.

\subsection{Experiment procedure}
\label{subsec:methods:experiment-procedure}

To maintain experimental control, participants were tested individually. Each participant executed the assigned tasks in the same environment using a preconfigured Windows 11 laptop and an Android 14 smartphone. In both cases, the moderator assigned the username to avoid collisions between credentials on each tested end device. Participants were also instructed to use passwords or passkeys according to their assigned group. Passwords were chosen by participants but had to meet minimum NIST requirements~\cite{temoshok_digital_2025} (at least 8 characters), consistent with Lyastani et al.~\cite{lyastani_is_2020}. Passkeys had to be registered on the platform authenticator of the end device (not on an external authenticator). To ensure consistency, both password and passkey methods require users to identify themselves with a username before authentication. Therefore, passkeys in this experiment are not based on the FIDO2 usernameless authentication flow.

The following procedure was followed for each participant, with adaptations according to the assigned authentication method:

\begin{itemize}
	\item \textbf{Stage 0: Preparation}. The captive portal system is reset and all previous registered users are removed from the IAM database. On each end device, the browsing history and cookies are deleted; and they are connected to the Wi-Fi network to allow connectivity during registration. Finally, both devices are screen-locked.
	\item \textbf{Stage 1: Pre-experiment}. The user signs the informed consent form and fills in the initial demographics questionnaire. Once completed, the moderator explains that this study tests a Wi-Fi system and shows both end devices to the user. Additionally, the moderator states that the predefined PIN to unlock both devices is ``1212'', which should be used whenever necessary. For users in $G1_{\text{passkey}}$, a brief text about passkeys is read to balance different levels of prior knowledge about this authentication method.
	
	\item \textbf{Stage 2.1: Registration and authentication in Windows}. The user performs two tasks in the Windows laptop, using their assigned authentication method. Before each task, the user is required to unlock the laptop using the PIN.
	
	\begin{itemize}
		\item \textit{TASK\_WIN1: Registration task}. The user is told to register using the ``Registration'' shortcut available on the desktop. Then, the registration portal requests the name and username, and displays two buttons: password and passkey. Users should choose their assigned authentication method, as indicated by the moderator. Passkeys are registered through the Windows Hello prompt, where the user is prompted to enter the PIN and confirm the operation.
		\item \textit{TASK\_WIN2: Authentication task}. Next, the user is instructed to connect to the Wi-Fi network using the registered credentials, without assistance. The user selects the network and clicks the connection button. The Windows Captive Portal Detection (CPD) system opens the default browser with the captive portal login UI, where the username should be entered. Users who have registered a passkey follow the Windows Hello prompt, confirming the operation with the PIN. Once authenticated, the user may test Internet connectivity.
	\end{itemize}
	
	\item \textbf{Stage 2.2: Registration and authentication in Android}. According to the experiment design, the user now performs registration and authentication tasks in the Android smartphone. Both tasks are similar to Windows. Before each task, the user must unlock the device using the PIN.
	
	\begin{itemize}
		\item \textit{TASK\_ANDR1: Registration task}. The user uses the shortcut on the Android home screen and fills in the required data in the registration form, selecting their assigned authentication method. Passkeys are registered through the Android OS prompt, where the user selects ``This device'' as the target destination for the passkey and completes the operation by entering the PIN.
		\item \textit{TASK\_ANDR2: Authentication task}. Using the Android UI, the user connects to the Wi-Fi network. The Android CPD system opens a captive portal mini-browser. It is important to note that nowadays this mini-browser is not compatible with the WebAuthn API. To overcome this restriction, our captive portal system implements an URL redirection mechanism that requests opening another browser, causing the Android OS to show a warning to the user. Once in a compatible browser, the user enters the username and follows the instructions in the login UI. Users who have registered a passkey follow the Android OS passkey prompt, confirming the operation with the PIN. Once authenticated, the user may test Internet connectivity.
	\end{itemize}

	\item \textbf{Stage 3: Retention test}. To study the learnability of the authentication task, the user is asked to repeat \textit{TASK\_WIN2} and \textit{TASK\_ANDR2}. The execution of these tasks is expected to be faster than during the first exposure to the authentication system.

	\item \textbf{Stage 4: Post-experiment}. After the execution of all tasks and the retention test, the user is asked to fill in a standard System Usability Scale (SUS) questionnaire about their overall experience using the Wi-Fi authentication system.

\end{itemize}

During the execution of the experiment, data collection is based on screen and audio recordings and several surveys. To counterbalance learning effects in the split-plot design, the order of exposure to the devices was randomly assigned to participants, resulting in half of the participants of both groups performing the Windows tasks first, and the other half performing the Android tasks first. This design helped mitigating possible bias from learning effects of users when performing tasks in the second device. The data collection methods are detailed in the following sections.

\subsection{Quantitative data collection and analysis}
\label{subsec:methods:quantitative-data}
Measuring system usability, as defined by ISO 9241~\cite{noauthor_iso_nodate}, requires assessing (1) effectiveness, (2) efficiency, and (3) user satisfaction. In our lab study, we collected quantitative data for these three measures using observational data and post-task and post-experiment surveys.

\paragraph{OBSERVATION} To measure task effectiveness and efficiency~\cite{lazar_research_2017} for registration and authentication, we observed both tasks on each operating system. For this purpose, we recorded the screen and audio during each task, stopping the recording once the task was completed or abandoned. For each participant, we recorded six videos: two for the registration tasks (one per operating system) and four for the two authentication tasks (one per operating system) and their corresponding retention tests. Additionally, these videos were complemented with observational notes gathered by the researcher conducting the laboratory experiment.

These recordings and notes allowed us to measure the task completion rate, time-on-task, and error rate for each task and participant. While time-on-task was directly measured using the video length, the completion rate and error rates required manual coding of the observational data, yielding a table with the following variables for each participant:

\begin{itemize}
	\item \textbf{Registration completion} (Windows and Android). Indicates whether the participant successfully completed registration and created a valid account.
	\item \textbf{Authentication completion} (Windows and Android, two attempts each). Indicates whether the participant completed authentication with the assigned method (passkey or password) within the captive portal and reached a success message.
	\item \textbf{Internet connection completion} (Windows and Android, two attempts each). Indicates whether the participant accessed web content on the Internet after authentication.
	\item \textbf{Task duration} (for each registration and authentication task). Measured in seconds; the time spent completing or abandoning the task.
	\item \textbf{Errors (by type)}. This field includes one variable per error type, representing the number of errors that occurred during task execution. Here, we considered errors caused by the system and the user, which interrupted or conditioned the normal execution of the task. Error types were categorised and encoded manually by a researcher using the experiment notes and the audio/video recordings for each task.
\end{itemize}

\paragraph{SURVEYS} User satisfaction was measured using Likert-scale questions in post-task and post-experiment surveys~\cite{barnum_usability_2021}. We conducted a total of three surveys for each participant: two after the first authentication task on each operating system and one after the experiment. The complete questionnaires are included in~\ref{app:post-task-survey} and~\ref{app:post-experiment-survey}.

\begin{itemize}
	\item \textbf{Post-task surveys}. They included five 5-point Likert-scale questions and the standard 7-point Likert-scale Single Ease Question (SEQ). With these surveys, we captured participants' satisfaction and perceived usability for authentication on each operating system.
	\item \textbf{Post-experiment survey}. After the experiment, we asked participants to fill in the standard System Usability Scale (SUS) questionnaire with ten 5-point Likert-scale questions~\cite{brooke_sus_1996,ruoti_authentication_2015,borsci_assessing_2015}. To measure the overall perceived usability, we calculated the corresponding SUS scores for each participant (see Equation~\ref{eq:sus}).
\end{itemize}

\begin{equation}
	\label{eq:sus}
	\text{SUS} = 2.5 \left( \sum_{i \in \{1,3,5,7,9\}} (Q_i - 1) + \sum_{i \in \{2,4,6,8,10\}} (5 - Q_i) \right)
\end{equation}

\subsection{Qualitative data collection and analysis}
\label{subsec:methods:qualitative-data}
Post-task and post-experiment surveys also included open-ended questions to complete the measurement of user perception, satisfaction and acceptance:

\begin{itemize}
	\item \textbf{Post-task surveys}. In post-task surveys, we collected suggestions from participants for both registration and authentication in each operating system (see \ref{app:post-task-survey}).
	\item \textbf{Post-experiment survey}. At the end of the experiment, we asked participants to describe their general experience with the system and to indicate its advantages and disadvantages (see \ref{app:post-experiment-survey}).
\end{itemize}

These qualitative data were analysed using codebooks. To build these codebooks, two independent researchers created codebooks in parallel by analysing all participant responses for each question. Once these codebooks were created, they were merged and refined. The final codebooks are included in \ref{appendix:codebooks}, together with the description, their frequency and examples. These codebooks were used to encode every answer to these open-ended questions, allowing us to perform an analysis of the differences between the perceived usability and the participant acceptance among the groups and operating systems.

\subsection{Participants and demographics}
We conducted our lab study with 50 participants between September and October 2025. We recruited them using chat groups related to the university, and during lessons in the faculty of computer science. Interested participants self-registered in an online form, where they were informed about the experiment, giving their consent to participate. Later, they had to book the in-person meeting for the individual laboratory study, from a large set of free time slots during more than one week. Participants were compensated by participating in a raffle to win one of the 4 surprise prizes, valued at more than \euro25.

\begin{figure}[ht!]
	\centering
	\includegraphics[width=\linewidth]{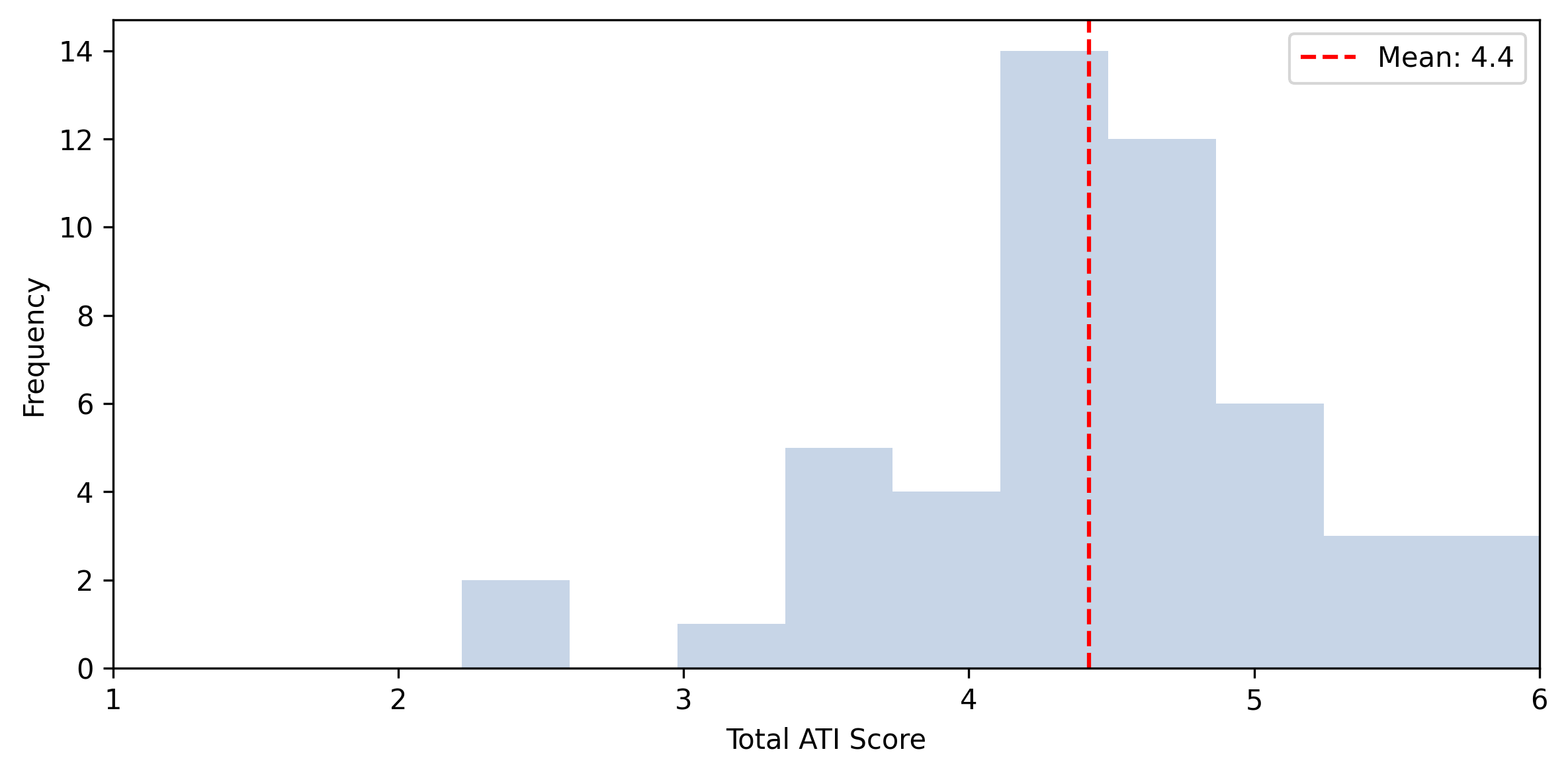}
	\caption{Histogram of the Affinity for Technology Interaction (ATI) questionnaire among participants.}
	\label{fig:methods:ati-histogram}
\end{figure}

Before performing the tasks in the laboratory, participants completed a demographics questionnaire. Of the sample of participants (N=50), 40 (80\%) were identified as male and 10 (20\%) were identified as female. Most participants (48) were enrolled in IT-related degree programmes at the university, and selected the age range of 18 - 25 years old. Two participants selected the range of 26 - 35 years old, one of which was studying a MSc and the other was a PhD candidate related to other scientific fields. All participants responded to a standard Affinity for Technology Interaction (ATI) questionnaire~\cite{franke_personal_2019}, using 6-point Likert items, which resulted in a mean affinity of 4.42 ($\pm 0.74$), as shown in Figure~\ref{fig:methods:ati-histogram}.

\begin{figure}[ht!]
	\centering
	\includegraphics[width=\linewidth]{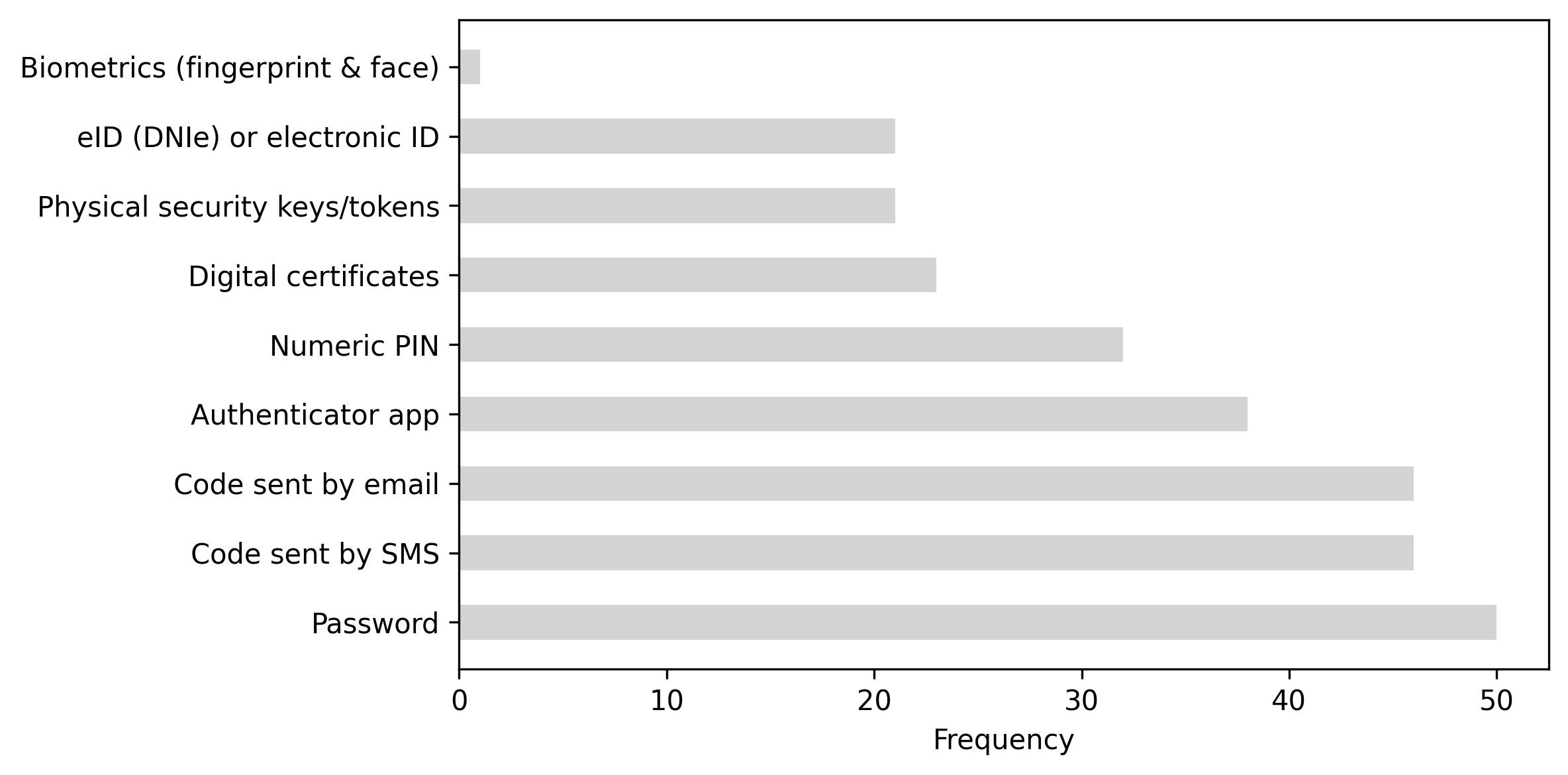}
	\caption{Most common authentication methods used by participants in their online accounts.}
	\label{fig:methods:auth-methods-bar}
\end{figure}

\begin{figure}[ht!]
	\centering
	\includegraphics[width=\linewidth]{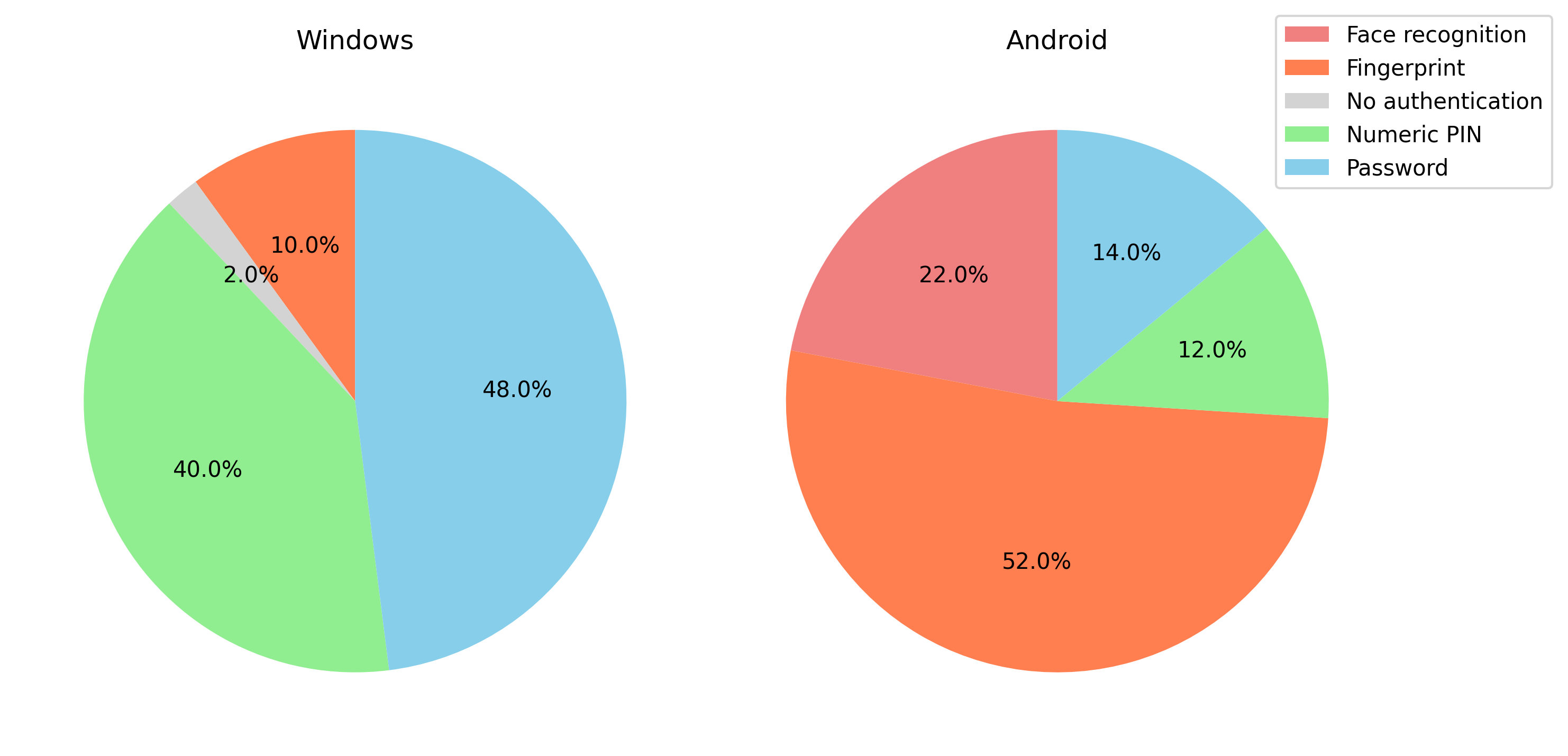}
	\caption{Most common authentication methods used by participants to lock their devices.}
	\label{fig:methods:locking-methods-pie}
\end{figure}

Most participants frequently use Windows (72\%) and Android (78\%), so they are familiar with the operating systems that are used in the usability experiment. In addition, most participants use more authentication methods beyond passwords in their online accounts, as shown in Figure~\ref{fig:methods:auth-methods-bar}. It is worth highlighting that 42\% of respondents claim to have used passkeys before. On the other hand, we have asked which authentication method participants used to lock their laptop and mobile phone. Results are shown in Figure~\ref{fig:methods:locking-methods-pie}. While the use of a PIN or a password is quite common when locking computers (88\%), in smartphones biometric authentication is the most common (74\%) method.

\section{Quantitative results}
\label{sec:quantitative-results}
To describe the usability of each authentication method in this split-plot study, we present results on task effectiveness, efficiency, and user satisfaction after completing the registration and authentication tasks. Post-task questionnaires, observational data collection, and the final usability questionnaire produced the quantitative results presented in this section.

\subsection{Effectiveness and efficiency}

The task effectiveness and efficiency were measured using the observational data collected during the experiment. To compare both authentication methods, we measured the task completion ratio, the error rates and time on task.

\paragraph{\small COMPLETION RATIO}

\begin{figure*}[ht!]
	\centering
	\includegraphics[width=\linewidth]{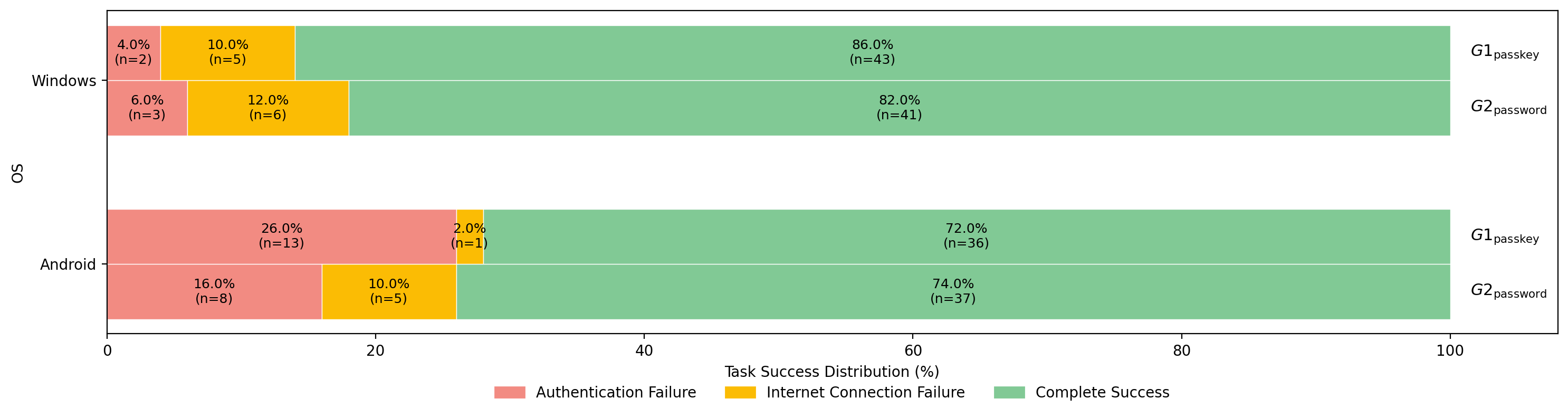}
	\caption{Task completion of the authentication task in the captive portal, divided by group and operating system.}
	\label{fig:results:task-success-bars}
\end{figure*}

Figure~\ref{fig:results:task-success-bars} shows completion of the captive portal authentication task, grouped by operating system and authentication method. Completing the task includes both successful authentication and testing Internet connectivity. Results show very similar completion ratios for both authentication methods. When users connected to the captive portal network using Windows, passkeys yielded an 86\% success rate, while passwords yielded 82\% ($\chi^2 (1, N=100)=0.07,\ p=0.785$). Task completion when using Android is lower, but also similar for both authentication methods: 72\% success when authenticating using passkeys, and 74\% when using passwords ($\chi^2 (1, N=100)=0.00,\ p=1.000$).

Tables~\ref{table:results:registration-completion} and~\ref{table:results:authentication-completion} show the registration and authentication completion rates, respectively. While registration achieved 100\% completion in all cases, authentication completion differs across operating systems, authentication methods, and connection attempts. To measure the learnability of the authentication task, users executed the authentication task twice for each operating system: a first attempt and a retention test at the end of the experiment. Table~\ref{table:results:authentication-completion} indicates a higher completion rate in the retention test when using passkeys, while passwords show a more consistent completion rate across attempts. First-to-second attempt completion improved for passkeys from 72.0\% to 86.0\% ($\Delta=+14.0$ pp; $\chi^2(1, N=100)=2.17,\ p=0.141$), whereas passwords were stable from 78.0\% to 78.0\% ($\Delta=+0.0$ pp; $\chi^2(1, N=100)=0.00,\ p=1.000$).

\begin{table*}[ht!]
	\centering
	\resizebox{0.75\textwidth}{!}{%
		\begin{tabular}{|l|l|l|l|l|l|l|}
			\hline
			\textbf{Task}        & \textbf{$G1_{\text{passkey}}$ \small{Total}} & \textbf{$G1_{\text{passkey}}$ (\%)} & \textbf{$G2_{\text{password}}$ \small{Total}} & \textbf{$G2_{\text{password}}$ (\%)} & \textbf{\small{Overall Total}} & \textbf{\small{Overall (\%)}} \\ \hline
			Windows Registration & 25                   & 100                   & 25                    & 100                    & 50                   & 100                   \\ \hline
			Android Registration & 25                   & 100                   & 25                    & 100                    & 50                   & 100                   \\ \hline
		\end{tabular}%
	}
	\caption{Task completion of the registration task for each operating system and group.}
	\label{table:results:registration-completion}
\end{table*}

\begin{table*}[!ht]
	\centering
	\resizebox{0.7\textwidth}{!}{%
	\begin{tabular}{|l|l|l|l|l|l|l|l|l|}
		\hline
		\textbf{Task} &
		\textbf{Group} &
		\textbf{Attempt} &
		\textbf{\begin{tabular}[c]{@{}l@{}}Auth \\ Total\end{tabular}} &
		\textbf{\begin{tabular}[c]{@{}l@{}}Auth \\ (\%)\end{tabular}} &
		\textbf{\begin{tabular}[c]{@{}l@{}}Internet \\ Total\end{tabular}} &
		\textbf{\begin{tabular}[c]{@{}l@{}}Internet \\ (\%)\end{tabular}} &
		\textbf{\begin{tabular}[c]{@{}l@{}}Complete Task \\ Total\end{tabular}} &
		\textbf{\begin{tabular}[c]{@{}l@{}}Complete Task \\ (\%)\end{tabular}} \\ \hline
		\multirow{4}{*}{Windows Authentication} & \multirow{2}{*}{$G1_{\text{passkey}}$}  & \begin{tabular}[c]{@{}l@{}}First\end{tabular} & 24 & 96 & 19 & 76 & 19 & 76 \\ \cline{3-9} 
		&                           & \begin{tabular}[c]{@{}l@{}}Second\end{tabular} & 24 & 96 & 24 & 96 & 24 & 96 \\ \cline{2-9} 
		& \multirow{2}{*}{$G2_{\text{password}}$} & \begin{tabular}[c]{@{}l@{}}First\end{tabular} & 24 & 96 & 21 & 84 & 21 & 84 \\ \cline{3-9} 
		&                           & \begin{tabular}[c]{@{}l@{}}Second\end{tabular} & 23 & 92 & 20 & 80 & 20 & 80 \\ \hline
		\multirow{4}{*}{Android Authentication} & \multirow{2}{*}{$G1_{\text{passkey}}$}  & \begin{tabular}[c]{@{}l@{}}First\end{tabular} & 17 & 68 & 24 & 96 & 17 & 68 \\ \cline{3-9} 
		&                           & \begin{tabular}[c]{@{}l@{}}Second\end{tabular} & 20 & 80 & 23 & 92 & 19 & 76 \\ \cline{2-9} 
		& \multirow{2}{*}{$G2_{\text{password}}$} & \begin{tabular}[c]{@{}l@{}}First\end{tabular} & 22 & 88 & 18 & 72 & 18 & 72 \\ \cline{3-9} 
		&                           & \begin{tabular}[c]{@{}l@{}}Second\end{tabular} & 20 & 80 & 21 & 84 & 19 & 76 \\ \hline
	\end{tabular}%
	}
	\caption{Task completion of the authentication task for each operating system and group. Completion rates included for the first attempt; and for the second attempt, the retention test.}
	\label{table:results:authentication-completion}
\end{table*}

\paragraph{\small TASK ABANDONMENT}
Users abandoned the task at two points: (1) abandoning before completing authentication; and (2) completing authentication but not achieving Internet connectivity and therefore abandoning the task. No user abandoned the task at the first point. When users started the task, they were required to connect to the Wi-Fi network and follow operating system instructions to open a compatible web browser for authentication; these instructions differ by operating system. After authenticating successfully, users were asked to check whether Internet access was available. Most users who detected that Internet access was not working after authentication repeated the task.

As shown in red in Figure~\ref{fig:results:task-success-bars}, abandoning without completing authentication was most frequent for passkeys on Android, followed by passwords on Android. In the same figure (orange), abandoning after detecting an Internet connectivity failure following successful authentication was more frequent on Windows for both methods, and on Android for passwords. Most of these cases involved more than one task repetition.

\begin{figure}[ht!]
	\centering
	\includegraphics[width=0.95\linewidth]{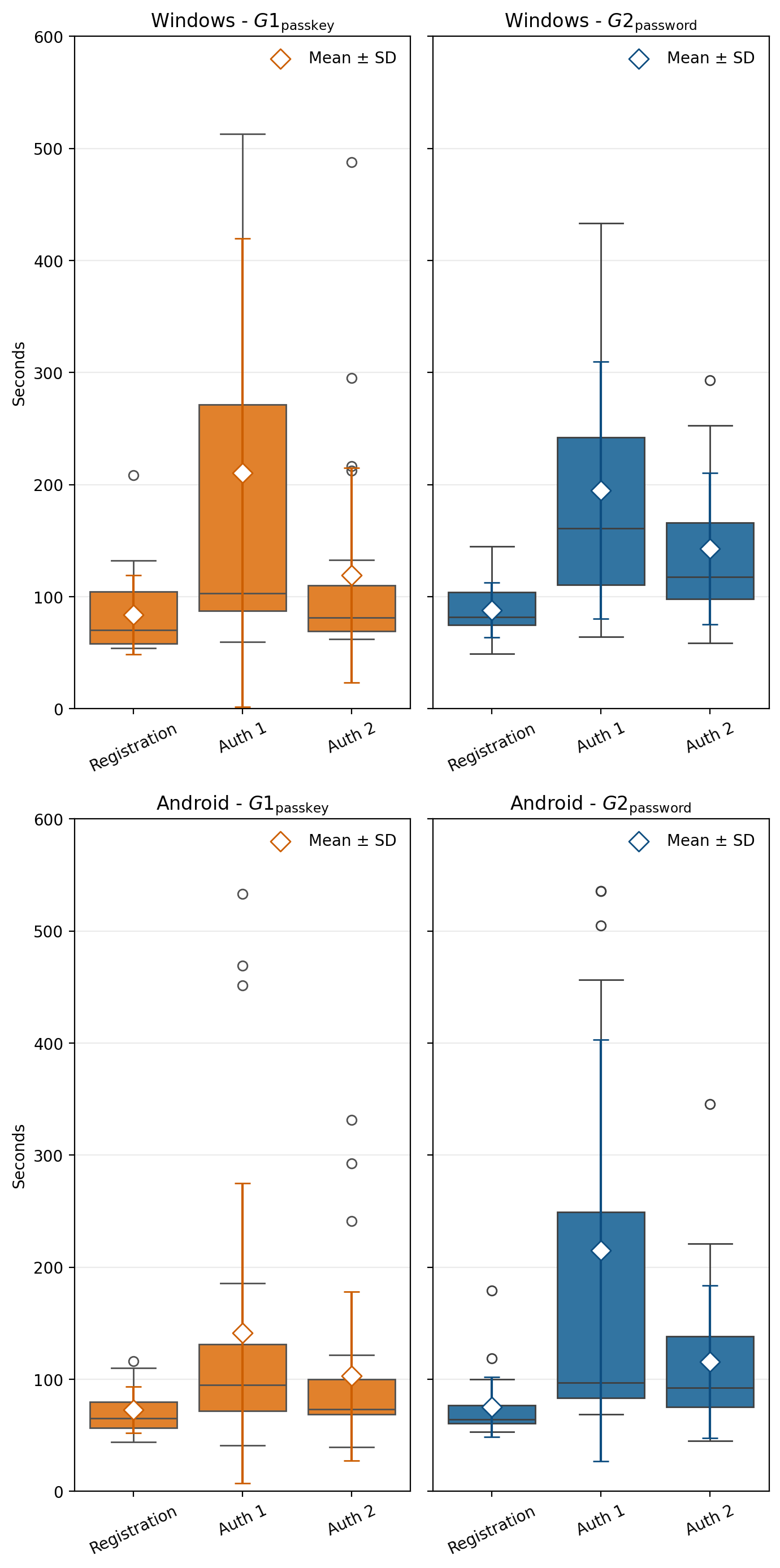}
	\caption{Efficiency of registration and authentication: time-on-task. In each scenario, the figure shows the elapsed time for registration and for the two authentication attempts. Times were measured until completion or abandonment by users. To improve readability, two outliers above 700 seconds are not shown in the plot but are included in the calculations: an Android password case (P34) with 715~s and a Windows passkey case (P31) with 990~s.}
	\label{fig:results:time-on-task-boxplot}
\end{figure}

\paragraph{\small TIME ON TASK}
Because some users had to repeat the task due to various errors, task efficiency shows high variability. Figure~\ref{fig:results:time-on-task-boxplot} shows time-on-task (seconds) for each operating system and authentication method, with variability represented via box plots. To illustrate differences between the first connection and the second connection (retention test), the figure shows the measured elapsed time for each task attempt separately. As expected, in every scenario the mean time of the second authentication attempt was lower than that of the first attempt, indicating retention effects.

Table~\ref{table:results:time-on-task} reports time-on-task (seconds) for registration and authentication across operating systems, authentication methods, and attempts. Registration times are similar between methods: on Windows, the mean registration time for passkeys was 83.86~s (median 70.50~s, SD 35.22~s) versus 87.97~s (median 82.10~s, SD 24.44~s). On Android, registration times for both methods are also similar, but time-on-task is approximately 10~s lower than on Windows, with a mean of 72.83~s for passkeys and 75.32~s for passwords. Despite differences between operating systems, the close medians and means, together with similar standard deviations for the two authentication methods, indicate consistent registration performance for passwords and passkeys on both platforms. Specifically, Mann-Whitney U-Tests show that Windows registration time-on-task ($U=243.0,\ p=0.181$) and Android registration time-on-task ($U=294.0,\ p=0.727$) are not significantly different between methods. However, authentication time-on-task shows clearer differences. On Windows, passkey authentication is represented by a mean time of 163.9~s (median 96.0~s), compared with a mean of 169.0~s (median 136.4~s) for passwords ($U=897.5,\ p=0.022$). On Android, passkey authentication yielded a mean time of 122.5~s (median 79.9~s), compared with a mean of 165.2~s (median 94.4~s) for passwords ($U=925.5,\ p=0.036$).

\paragraph{\small LEARNING EFFECT}
Authentication shows a pronounced learning effect and substantial heterogeneity in first attempts, as shown by Wilcoxon signed-rank tests. On Windows, passkey authentication time fell from 210.54~s on the first attempt (median 108.42~s, SD 209.02~s) to 121.30~s (median 81.57~s, SD 95.66~s), a mean reduction of 89.24~s ($W=240.0,\ p=0.004$); whereas password authentication fell from 194.96~s (median 161.20~s, SD 161.20~s) to 143.04~s (median 117.77~s, SD 67.53~s) ($W=237.0,\ p=0.023$). On Android, passkey means decreased from 141.23~s to 102.93~s (-38.30~s) ($W=203.0,\ p=0.068$) and password means from 215.03~s to 115.47~s (-99.56~s) ($W=269.0,\ p=0.002$). The large gaps between means and medians, together with the very high standard deviations on first attempts, indicate outliers with long completion times; in the second attempt, both methods converge to lower authentication times with a less dispersed distribution. Overall, while registration is comparable across methods, authentication is characterised by high initial variability and clear improvement on repeat attempts, with passkeys showing consistently lower medians and a greater reduction in dispersion after the first attempt.

Considering the authentication completion rate in each attempt, we observe a clear learning effect for the $G1_{\text{passkey}}$ group on Windows. While the first authentication attempt with passkeys on Windows was completed by only $76\%$ of participants, the second attempt reached $96\%$ success ($\Delta=+20.0$ pp; $\chi^2(1, N=50)=2.66,\ p=0.103$). Results for $G2_{\text{password}}$ were different: $4\%$ of users who completed password authentication in the first attempt failed in the second attempt both for Windows and Android ($\chi^2(1, N=50)=0.00,\ p=1.000$).


\begin{table*}[!ht]
	\centering
	\resizebox{0.72\textwidth}{!}{%
		\begin{tabular}{|l|l|l|r|r|r|r|r|r|r|}
			\hline
			\textbf{Task} &
			\textbf{Group} &
			\textbf{Attempt} &
			\multicolumn{1}{l|}{\textbf{N}} &
			\multicolumn{1}{l|}{\textbf{Mean Time}} &
			\multicolumn{1}{l|}{\textbf{Median Time}} &
			\multicolumn{1}{l|}{\textbf{Std Time}} &
			\multicolumn{1}{l|}{\textbf{Q1}} &
			\multicolumn{1}{l|}{\textbf{Q3}} &
			\multicolumn{1}{l|}{\textbf{IQR}} \\ \hline
			\multirow{2}{*}{Windows Registration}   & $G1_{\text{passkey}}$                   & \multicolumn{1}{c|}{-} & 25 & 83.87  & 70.50  & 35.22  & 58.10  & 104.63 & 46.53  \\ \cline{2-10} 
				& $G2_{\text{password}}$                  & \multicolumn{1}{c|}{-} & 25 & 87.97  & 82.10  & 24.44  & 74.67  & 104.00 & 29.33  \\ \hline
			\multirow{2}{*}{Android Registration}   & $G1_{\text{passkey}}$                   & \multicolumn{1}{c|}{-} & 25 & 72.83  & 65.21  & 20.53  & 56.72  & 79.91  & 23.19  \\ \cline{2-10} 
				& $G2_{\text{password}}$                  & \multicolumn{1}{c|}{-} & 25 & 75.32  & 64.46  & 26.81  & 60.58  & 77.04  & 16.46  \\ \hline
			\multirow{4}{*}{Windows Authentication} & \multirow{2}{*}{$G1_{\text{passkey}}$}  & First                  & 24 & 210.54 & 108.42 & 209.02 & 88.72  & 295.80 & 207.08 \\ \cline{3-10} 
				&                           & Second                 & 25 & 119.07 & 81.57  & 95.66  & 69.30  & 109.90 & 40.60  \\ \cline{2-10} 
				& \multirow{2}{*}{$G2_{\text{password}}$} & First                  & 25 & 194.96 & 161.20 & 114.67 & 110.67 & 242.43 & 131.76 \\ \cline{3-10} 
				&                           & Second                 & 25 & 143.04 & 117.77 & 67.53  & 98.20  & 166.30 & 68.10  \\ \hline
			\multirow{4}{*}{Android Authentication} & \multirow{2}{*}{$G1_{\text{passkey}}$}  & First                  & 25 & 141.23 & 94.94  & 133.90 & 71.93  & 131.28 & 59.35  \\ \cline{3-10} 
				&                           & Second                 & 24 & 102.93 & 73.18  & 75.32  & 68.90  & 99.96  & 31.06  \\ \cline{2-10} 
				& \multirow{2}{*}{$G2_{\text{password}}$} & First                  & 25 & 215.03 & 98.45  & 187.91 & 84.12  & 276.62 & 192.50 \\ \cline{3-10} 
				&                           & Second                 & 25 & 115.47 & 92.27  & 68.02  & 75.20  & 138.43 & 63.23  \\ \hline
		\end{tabular}%
	}
	\caption{Time-on-task (seconds) for each operating system and group.}
	\label{table:results:time-on-task}
\end{table*}

\paragraph{\small ERROR RATE}
The overall error rate during task execution was 49.68\%, with slightly more errors when using passwords than passkeys. The most pronounced differences appeared across operating systems: Windows showed an error rate of 35.71\%, whereas Android reached 65.75\%. This difference was statistically significant ($\chi^2(1, N=157)=12.92,\ p<0.001$). However, when comparing the error rates between authentication methods, the $G1_{\text{passkey}}$ group had an error rate of 45.57\%, while the $G2_{\text{password}}$ group had an error rate of 53.85\%, a difference that was not statistically significant ($\chi^2(1, N=157)=0.77,\ p=0.380$).

Table~\ref{table:results:errors} summarises the observed error types, ordered by frequency. Three of these errors stem from the Wi-Fi infrastructure, while the remainder originate from user actions, such as incorrect username/password input or unintended changes to network settings.

The most frequent error is not user-related but caused by connectivity issues in the captive portal during the interval between user authentication and client authorisation. Because most users test Internet access immediately after authenticating, they often experience failures if authorisation has not yet completed (see Section~\ref{sec:system-overview}). 

Additional user-originated errors arise from misleading user interface elements, such as the \emph{Manage Account} button displayed after authentication. On Android, the captive portal opens a browser compatible with the WebAuthn API, which forces an exit from the operating system’s Captive Portal Detection mini-browser and triggers a premature \emph{connected} notification, frequently confusing participants. \ref{appendix:observed-errors} includes a detailed description of these UI-originated errors, together with sample screenshots of the recordings during the execution of the experiment

Finally, some errors resulted from software bugs. The most common occurred when users double-clicked the button that initiates passkey registration or authentication. After the first click, the operating system opens a dialog over the browser to register or authenticate with a passkey. Occasionally, the operating system took about 1~s to display this dialog, leading users to click the button again and causing a WebAuthn signature verification failure on the server, invalidating authentication. This issue can be mitigated by disabling the button after the first click and re-enabling it on failure or timeout.

\begin{table*}[!ht]
	\centering
	\resizebox{\textwidth}{!}{%
		\begin{tabular}{|l|l|l|l|l|l|l|l|}
			\hline
			\textbf{Error Type} &
			\textbf{Description} &
			\textbf{Total} &
			\textbf{\begin{tabular}[c]{@{}l@{}}\%\end{tabular}} &
			\textbf{\begin{tabular}[c]{@{}l@{}}$G1_{\text{passkey}}$ \\ \small{Total}\end{tabular}} &
			\textbf{\begin{tabular}[c]{@{}l@{}}$G1_{\text{passkey}}$ \\ \small{(\%)}\end{tabular}} &
			\textbf{\begin{tabular}[c]{@{}l@{}}$G2_{\text{password}}$ \\ \small{Total}\end{tabular}} &
			\textbf{\begin{tabular}[c]{@{}l@{}}$G2_{\text{password}}$\\ \small{(\%)}\end{tabular}} \\ \hline
			Login: Captive Portal Internet Failure &
			\begin{tabular}[c]{@{}l@{}}After successful authentication, the user tests Internet connectivity \\ and it fails. The captive portal did not grant access yet.\end{tabular} &
			35 &
			70 &
			15 &
			60 &
			20 &
			80 \\ \hline
			Login: Manage Account Click &
			\begin{tabular}[c]{@{}l@{}}The user clicks on the ``Manage Account'' button after authentication,\\ which was not part of the task and leads to confusion.\end{tabular} &
			28 &
			56 &
			15 &
			60 &
			13 &
			52 \\ \hline
			Login: OS Notification Mislead &
			\begin{tabular}[c]{@{}l@{}}(Android) An OS push notification sent to the user with the message\\ ``Connected'' before authentication confuses the user.\end{tabular} &
			16 &
			32 &
			10 &
			40 &
			6 &
			24 \\ \hline
			Login: Username Input Error &
			During authentication, the user enters an incorrect username. &
			9 &
			18 &
			1 &
			4 &
			8 &
			32 \\ \hline
			Login: Forget Network &
			\begin{tabular}[c]{@{}l@{}}The user navigates to the OS settings and deletes (forgets) the experiment\\ Wi-Fi network, leading to a connectivity error in subsequent trials.\end{tabular} &
			7 &
			14 &
			2 &
			8 &
			5 &
			20 \\ \hline
			Login: Password Input Error &
			During authentication, the user enters an incorrect password. &
			6 &
			12 &
			0 &
			0 &
			6 &
			24 \\ \hline
			Login: Passkey Double Click &
			\begin{tabular}[c]{@{}l@{}}During authentication, the user presses the authentication button twice,\\ leading to a passkey authentication error due to a bug in the software.\end{tabular} &
			6 &
			12 &
			6 &
			24 &
			0 &
			0 \\ \hline
			Registration: Username Input Error &
			During registration, the user mistypes their username. &
			2 &
			4 &
			0 &
			0 &
			2 &
			8 \\ \hline
			Registration: Passkey Double Click &
			\begin{tabular}[c]{@{}l@{}}During registration, the user presses the registration button twice,\\ leading to a passkey registration error due to a bug in the software.\end{tabular} &
			2 &
			4 &
			2 &
			8 &
			0 &
			0 \\ \hline
			Login: Exit Captive Portal &
			\begin{tabular}[c]{@{}l@{}}The user closes the browser with the captive portal page without\\ completing authentication.\end{tabular} &
			1 &
			2 &
			0 &
			0 &
			1 &
			4 \\ \hline
			Login: WiFi Connection Error &
			\begin{tabular}[c]{@{}l@{}}The OS displays an error in the Wi-Fi settings with the message\\ ``not possible to connect to this network'', confusing the user.\end{tabular} &
			1 &
			2 &
			1 &
			4 &
			0 &
			0 \\ \hline
			Login: Back Button Press &
			The user presses the ``back'' button before completing authentication. &
			1 &
			2 &
			0 &
			0 &
			1 &
			4 \\ \hline
		\end{tabular}%
	}
	\caption{Efficiency during task execution: most common errors during registration and authentication.}
	\label{table:results:errors}
\end{table*}

\subsection{User satisfaction}
Observational measures were used to assess task effectiveness and efficiency. However, to assess system usability, it is also important to measure user perception, i.e. satisfaction after completing the tasks. For this reason, we conducted a post-task questionnaire after participants completed authentication with the assigned method. This questionnaire included the standard Single Ease Question (SEQ) and other Likert-scale questions to measure different aspects of perceived usability and security of authentication in the captive portal. Additionally, at the end of the experiment, we measured overall perceived usability using the post-experiment System Usability Scale (SUS) questionnaire.

\begin{figure*}[ht!]
	\centering
	\includegraphics[width=0.65\linewidth]{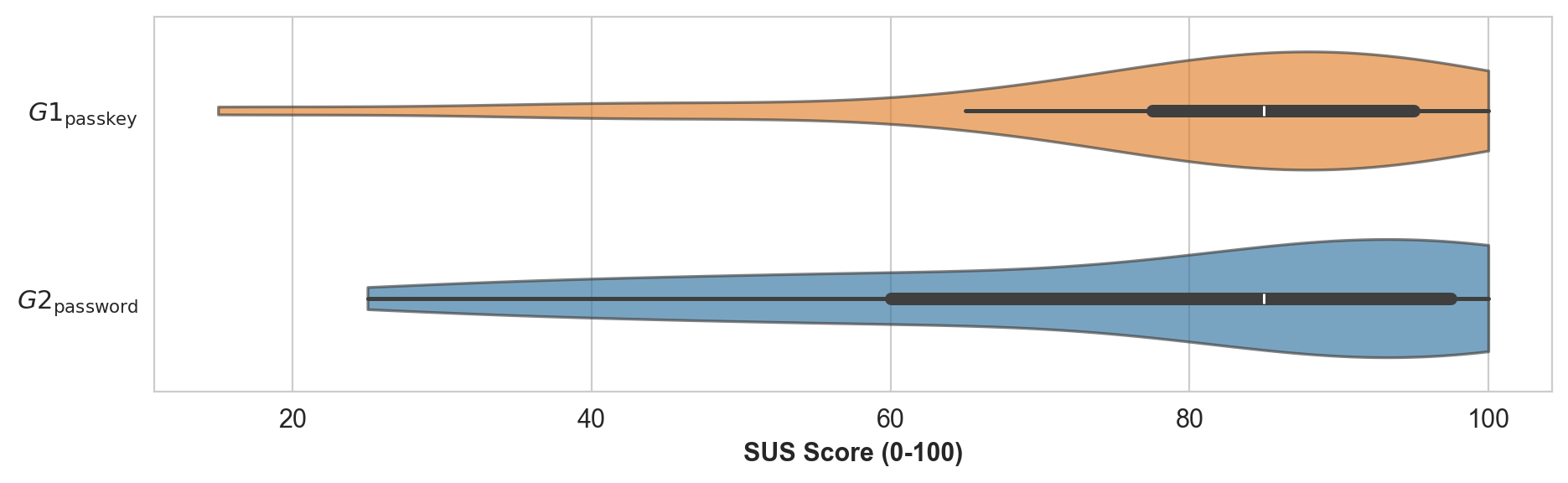}
	\caption{System Usability Scale (SUS) total score distribution of the perceived usability of the system for both groups after completing the experiment.}
	\label{fig:results:sus-boxplot}
\end{figure*}

Figure~\ref{fig:results:sus-boxplot} shows the distribution of SUS scores for each group of participants. Perceived system usability differs slightly across authentication methods: the mean SUS score is 80.0 for passkeys and 77.7 for passwords ($U=302.50,\ p=0.850,\ r=-0.03$). However, the median SUS value is 85.0 for both authentication methods, which is commonly interpreted as ``good usability'' in industry practice.

\begin{figure*}[ht!]
	\centering
	\includegraphics[width=0.8\linewidth]{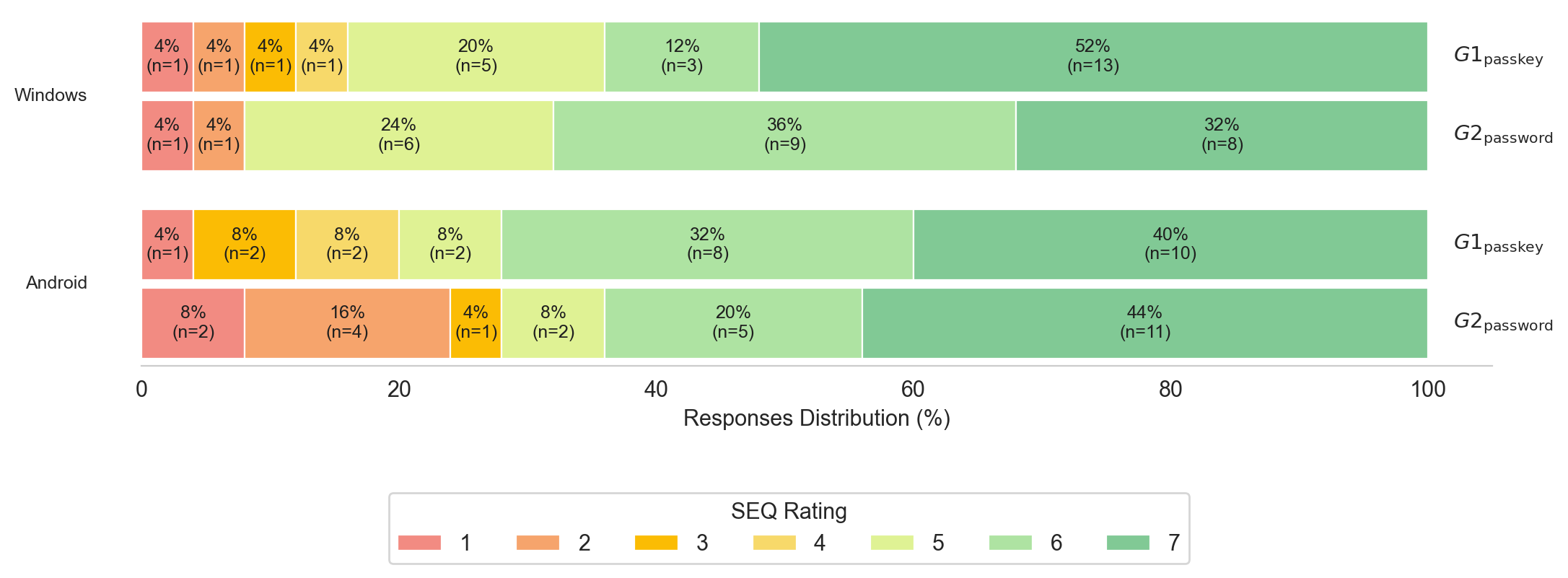}
	\caption{Single Ease Question (SEQ) response distribution after completing the authentication task in each operating system for both groups.}
	\label{fig:results:seq-stacked}
\end{figure*}

To examine quantitative differences in perceived usability across operating systems, Figure~\ref{fig:results:seq-stacked} shows the percentage of discrete responses to the SEQ question in the post-task questionnaires. Overall, both authentication methods have a similar SEQ score, with a mean difference of 0.28 ($U=1286.00,\ p=0.800$;\ $r=0.03$), and perceived usability differs slightly between the two operating systems. In Windows, more than half of the participants using passkeys (52\%) reported the highest usability score (SEQ = 7), while only 32\% of participants using passwords achieved this rating ($\chi^2(1, N=50)=4.153,\ p=0.0416,\ \varphi=0.288$). In Android, the results for passwords show a substantial percentage of responses indicating low usability (SEQ $<$ 3) compared to passkeys ($\chi^2(1, N=50)=4.153,\ p=0.0416,\ \varphi=0.288$), although both authentication methods achieve similar proportions of high-usability ratings.

\begin{figure*}[ht!]
	\centering
	\includegraphics[width=\linewidth]{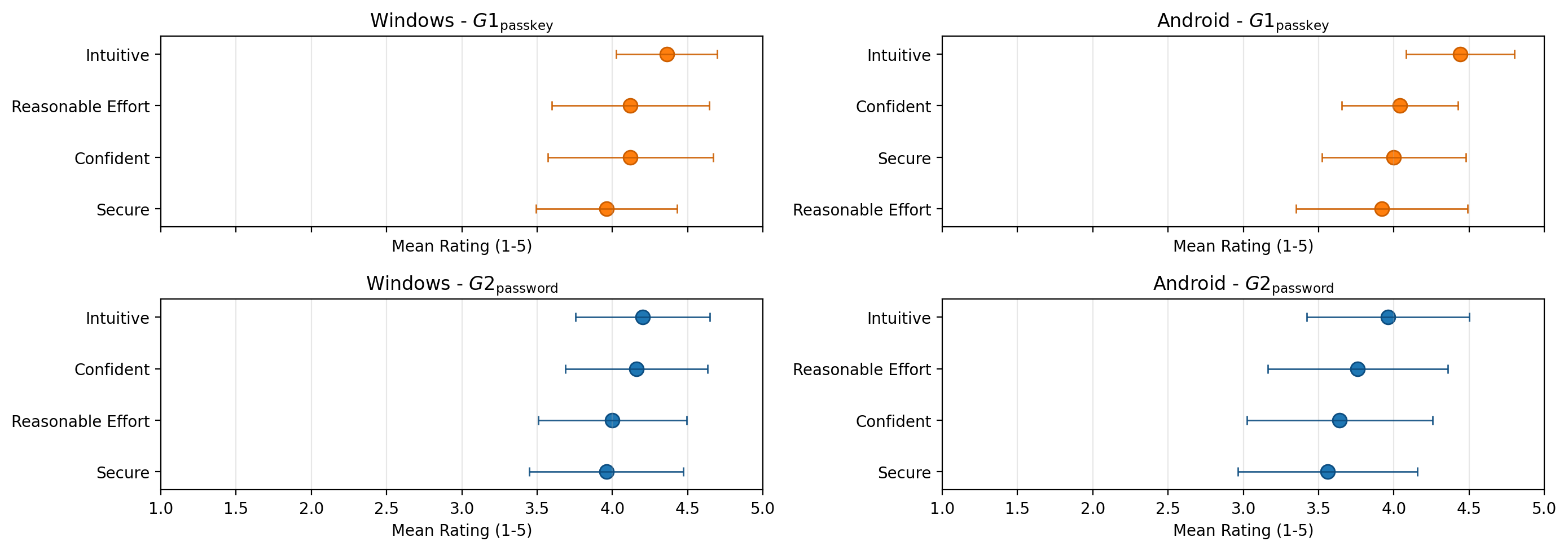}
	\caption{Mean values with confidence intervals of the responses to the Likert-scale satisfaction questions after the authentication task.}
	\label{fig:results:likert-scale-satisfaction-boxplot}
\end{figure*}

Together with the SEQ question, post-task questionnaires included four Likert-scale questions (see~\ref{app:post-task-survey}) to measure different aspects of the perceived usability and security of the system. Response statistics for each question across groups and operating systems are shown in Figure~\ref{fig:results:likert-scale-satisfaction-boxplot} and Table~\ref{table:results:likert-scale-satisfaction}. 

To compare responses to the four Likert-scale questions across groups and operating systems, we computed a composite score for each participant as the mean of the four responses, and then compared these values between groups for each operating system. On Windows, composite scores were very similar between $G1_{\text{passkey}}$ (mean 4.14 $\pm 0.83$) and $G2_{\text{password}}$ (mean 4.08 $\pm 0.88$), with no statistically significant difference ($U=321.00,\ p=0.876,\ r=0.027$). On Android, the gap was larger: $G1_{\text{passkey}}$ (mean 4.10 $\pm 0.78$) scored higher than $G2_{\text{password}}$ (mean 3.73 $\pm 1.24$). Although the Android metrics show a trend towards higher perceived usability scores when using passkeys, the difference is not statistically significant ($U=338.00,\ p=0.625,\ r=0.082$).

Across operating systems, perceived usability was generally high on Windows for both authentication methods. On Android, responses were more heterogeneous, with passwords showing lower composite scores and higher variability (median 4.00, IQR 2.00) than on Windows; passkey scores remained closer to the Windows distribution (median 4.50, IQR 0.75). These trends are consistent with Figure~\ref{fig:results:likert-scale-satisfaction-boxplot}, where the lowest per-question means are observed for passwords on Android.

\begin{table*}[!ht]
	\centering
	\resizebox{0.65\textwidth}{!}{%
		\begin{tabular}{|l|l|l|r|r|r|r|r|r|r|}
			\hline
			\textbf{Task} &
			\textbf{Group} &
			\textbf{Question} &
			\textbf{N} &
			\textbf{Mean} &
			\textbf{Median} &
			\textbf{Std} &
			\textbf{Q1} &
			\textbf{Q3} &
			\textbf{IQR} \\
			\hline
			
			\multirow{8}{*}{Windows Authentication} & \multirow{4}{*}{$G1_{\text{passkey}}$} & intuitive & 25 & 4.36 & 5.00 & 0.81 & 4.00 & 5.00 & 1.00 \\ \cline{3-10}
				& & confident & 25 & 4.12 & 5.00 & 1.33 & 4.00 & 5.00 & 1.00 \\ \cline{3-10}
				& & secure & 25 & 3.96 & 4.00 & 1.14 & 3.00 & 5.00 & 2.00 \\ \cline{3-10}
				& & reasonable\_effort & 25 & 4.12 & 5.00 & 1.27 & 4.00 & 5.00 & 1.00 \\ \cline{3-10}
			\cline{2-10}
				& \multirow{4}{*}{$G2_{\text{password}}$} & intuitive & 25 & 4.20 & 5.00 & 1.08 & 4.00 & 5.00 & 1.00 \\ \cline{3-10}
				& & confident & 25 & 4.16 & 5.00 & 1.14 & 4.00 & 5.00 & 1.00 \\ \cline{3-10}
				& & secure & 25 & 3.96 & 5.00 & 1.24 & 3.00 & 5.00 & 2.00 \\ \cline{3-10}
				& & reasonable\_effort & 25 & 4.00 & 4.00 & 1.19 & 4.00 & 5.00 & 1.00 \\
			\cline{2-10}
			\hline
			\multirow{8}{*}{Android Authentication} & \multirow{4}{*}{$G1_{\text{passkey}}$} & intuitive & 25 & 4.44 & 5.00 & 0.87 & 4.00 & 5.00 & 1.00 \\ \cline{3-10}
				& & confident & 25 & 4.04 & 4.00 & 0.93 & 4.00 & 5.00 & 1.00 \\ \cline{3-10}
				& & secure & 25 & 4.00 & 4.00 & 1.15 & 3.00 & 5.00 & 2.00 \\ \cline{3-10}
				& & reasonable\_effort & 25 & 3.92 & 5.00 & 1.38 & 3.00 & 5.00 & 2.00 \\ \cline{3-10}
			\cline{2-10}
				& \multirow{4}{*}{$G2_{\text{password}}$} & intuitive & 25 & 3.96 & 5.00 & 1.31 & 3.00 & 5.00 & 2.00 \\ \cline{3-10}
				& & confident & 25 & 3.64 & 4.00 & 1.50 & 2.00 & 5.00 & 3.00 \\ \cline{3-10}
				& & secure & 25 & 3.56 & 4.00 & 1.45 & 2.00 & 5.00 & 3.00 \\ \cline{3-10}
				& & reasonable\_effort & 25 & 3.76 & 4.00 & 1.45 & 2.00 & 5.00 & 3.00 \\
			\cline{2-10}
			\hline
		\end{tabular}%
	}
	\caption{Results per task and authentication method of the Likert-scale satisfaction questions after the authentication task.}
	\label{table:results:likert-scale-satisfaction}
\end{table*}

\section{Qualitative results}

To assess user perception, we included open-ended questions in both the post-task and post-experiment surveys (see \ref{app:post-task-survey} and \ref{app:post-experiment-survey}). After completing the registration and authentication tasks on each operating system, participants were asked to provide suggestions regarding these processes in the post-task questionnaires. At the end of the experiment (including the retention tasks), participants completed a post-experiment survey in which they described their overall experience of connecting to the Wi-Fi network using the assigned authentication method and identified perceived advantages and disadvantages of the system. All responses were analysed using codebooks (see \ref{appendix:codebooks}).

\subsection{Post-experiment questionnaire}

According to the post-experiment questionnaire, $60\%$ of participants in $G1_{\text{passkey}}$ reported a positive overall perception of system usability, compared with $40\%$ in $G2_{\text{password}}$. For example, P49 stated that they were \textit{``absolutely delighted [...]''} and that the system \textit{``[...] is very convenient and I wish it were integrated in the university [...]''}, while others provided similarly positive remarks, such as P3, who wrote that the system is \textit{``Slightly technical because it's not the usual way to connect to the internet, but completely viable and easy to use''}. Likewise, positive general comments were more frequent in $G1_{\text{passkey}}$ ($48\%$) than in $G2_{\text{password}}$ ($28\%$).

However, participants using passkeys reported more Android-related issues than those using passwords ($32\%$ vs.\ $16\%$), such as P13, who replied that the system is \textit{``Very simple; [but] on mobile it can be confusing at the end of the login when connecting to the Wi-Fi network, because it sends you back to the login screen when you're already connected [...]''}. Similarly, participants using passkeys more frequently expressed confusion about the connection process ($28\%$ vs.\ $12\%$); this concern was also raised by participants using passwords, such as P30:

\begin{quote}
	\textit{``The authentication system appears to be poorly integrated with the Wi-Fi network, leading to confusion and making users unsure whether the problem is with their device or the network.''} (P30, $G2_{\text{password}}$)
\end{quote}

The most frequently perceived advantages were largely consistent across both groups. As shown in Table~\ref{tab:results:advantages}, $26\%$ of participants described the system as simple and easy to use, and the same proportion identified security as a key advantage. For instance, P10 ($G1_{\text{passkey}}$) mentioned that the system is \textit{``a more private network and it also has double security, since you first need to log in and then enter the laptop's PIN''}.

\begin{table}[!ht]
	\resizebox{0.8\linewidth}{!}{%
		\begin{tabular}{@{}lll@{}}
			\toprule
			\textbf{Advantage}                      & \textbf{$G1_{\text{passkey}}$ (\%)} & \textbf{$G2_{\text{password}}$ (\%)} \\ \midrule
			Easy / simple to use                    & 24.0               & 28.0               \\
			Secure                                  & 24.0               & 28.0               \\
			Allows accurate access control          & 20.0               & 24.0               \\
			Fast authentication                     & 16.0               & 20.0               \\
			Trustworthy                             & 8.0                & 16.0               \\
			Comfortable to use                      & 8.0                & 12.0               \\
			Familiar or intuitive                   & 4.0                & 8.0                \\
			No common Pre-Shared-Key       			& 4.0                & 8.0                \\
			Requires no configuration               & 8.0                & 4.0                \\
			Can benefit from secure biometric auth.	& 8.0                & 0.0                \\
			Suitable for enterprise use             & 8.0                & 0.0                \\
			Suitable for events                     & 8.0                & 0.0                \\
			No need to remember a password 			& 8.0                & 0.0                \\
			Attractive UI                           & 4.0                & 4.0                \\
			None                                    & 4.0                & 4.0                \\
			\bottomrule
		\end{tabular}%
	}
	\caption{Advantages of the system identified by participants of the two groups after the experiment.}
	\label{tab:results:advantages}
\end{table}

However, participants using passwords more often mentioned generic benefits, such as speed, perceived trustworthiness, and familiarity or comfort during the connection process. For example, P6 stated:

\begin{quote}
	\textit{``Nowadays, passwords are one of the most widely used methods among internet users. The Wi-Fi network authentication in this experiment is quite similar to most other applications and websites that require a registered account [...]. Personally, I think it's a good idea.''} (P6, $G2_{\text{password}}$)
\end{quote}

In contrast, participants using passkeys reported advantages specific to the authentication mechanism, including ease of configuration, suitability for event and enterprise settings, and the ability to leverage biometric authentication while avoiding the need to remember passwords:

\begin{quote}
	\textit{``The main advantage I see is security: being able to have a passkey that doesn't have to be a password, which is always vulnerable to attacks or someone finding it out, etc. A method like a fingerprint is much more secure, since it's personal and uniquely identifies you, and it also removes the hassle of having to remember multiple different passwords.''} (P5, $G1_{\text{passkey}}$)
\end{quote}

Participants were also asked to report perceived disadvantages of the system. As detailed in Table~\ref{tab:results:disadvantages}, a substantial proportion of participants reported no disadvantages ($32\%$ in $G1_{\text{passkey}}$ and $24\%$ in $G2_{\text{password}}$). Both groups identified the primary drawback as the friction caused by repeatedly authenticating to connect to the Wi-Fi network ($16\%$ in $G1_{\text{passkey}}$ and $12\%$ in $G2_{\text{password}}$). For example, P34 clearly stated:

\begin{quote}
	\textit{``Having to log in becomes complicated and tedious. It seems very inconvenient to me to have to enter a username and password just to connect to a Wi-Fi network, aside from how annoying it is to have to use a web interface for the whole process.'' (P34, $G2_{\text{password}}$)}
\end{quote}

This drawback was followed by connection errors and security-related concerns (each $8\%$ in $G1_{\text{passkey}}$ and $12\%$ in $G2_{\text{password}}$), such as P30: \textit{``It's cumbersome to connect, with failures in the authentication system when connecting, which can make the user lose patience [...]''}.

\begin{table}[!ht]
	\resizebox{0.8\linewidth}{!}{%
		\begin{tabular}{@{}lll@{}}
			\toprule
			\textbf{Disadvantage}                        & \textbf{$G1_{\text{passkey}}$ (\%)} & \textbf{$G2_{\text{password}}$ (\%)} \\ \midrule
			None                                         & 32.0               & 24.0               \\
			Friction caused by authentication repetition & 16.0               & 12.0               \\
			Errors during the connection                 & 8.0                & 12.0               \\
			Has security risks                           & 8.0                & 12.0               \\
			Requires technical knowledge                 & 8.0                & 8.0                \\
			Many steps to connect                        & 4.0                & 12.0               \\
			Unfamiliar                                   & 8.0                & 8.0                \\
			Anybody can join the network                 & 8.0                & 8.0                \\
			Feels insecure / untrustworthy               & 8.0                & 4.0                \\
			A single authentication method is insecure   & 8.0                & 4.0                \\
			Privacy of the personal data                 & 8.0                & 0.0                \\
			Confusing UI                                 & 0.0                & 8.0                \\
			Requires an account to connect               & 8.0                & 4.0                \\
			Slow                                         & 4.0                & 4.0                \\
			Confusing connection process                 & 4.0                & 0.0                \\
			Requires specific hardware                   & 4.0                & 0.0                \\
			Requires to remember a password              & 0.0                & 4.0                \\
			\bottomrule
		\end{tabular}%
	}
	\caption{Disadvantages of the system identified by participants of the two groups after the experiment.}
	\label{tab:results:disadvantages}
\end{table}

Usability-related issues, such as requiring technical knowledge or being unfamiliar, were reported at similar rates in both groups ($8\%$ each), whereas participants in $G2_{\text{password}}$ more frequently mentioned the large number of steps required to connect and a confusing UI (both $12\%$ and $8\%$, respectively). In contrast, participants in $G1_{\text{passkey}}$ more often raised concerns related to the privacy of personal data~\cite{ali_privacy_2019} and hardware requirements (each $4$--$8\%$), like P31:

\begin{quote}
	\textit{``You don’t see step by step what you’re doing or your data; it seems to me like you’re giving data and trust to an external party, when nowadays a strong, long password for Wi-Fi connectivity is, in my opinion, more than sufficient.''} (P31, $G1_{\text{passkey}}$)
\end{quote}

\subsection{Post-task questionnaires}

Users completed brief post-task surveys, in which they were asked for suggestions to improve registration and login after using each operating system to connect to the Wi-Fi network through the captive portal.

Registration tasks received fewer suggestions than the authentication process. Most suggestions requested more instructions or guidance, while others reflected confusion about the username field. Furthermore, two participants in $G1_{\text{passkey}}$ suggested improving robustness by adding a password as a fallback (e.g. P36 after authenticating on Android): \textit{``Be able to set a password even if you're going to use a passkey, in case you change phones or are using multiple phones at the same time''}.

Suggestions related to the authentication process are summarised in Table~\ref{tab:results:suggestions-login}, excluding comments not aimed at proposing improvements (see codebooks in \ref{appendix:codebooks}). Participants in both groups frequently reported uncertainty about whether they were successfully connected to the network, leading to requests for clearer confirmation pages ($32\%$ in $G1_{\text{passkey}}$ and $12\%$ in $G2_{\text{password}}$). In this regard, many participants asked for more instructions and guidance during the connection process.

\begin{quote}
	\textit{``Provide a clearer confirmation that the process has been completed successfully and that you can start browsing.''} (P16 on Windows, $G2_{\text{password}}$)
\end{quote}

Performance-related concerns were more common in the password group, with $12\%$ requesting faster connections, compared with $4\%$ in the passkey group. In addition, $12\%$ of participants using passwords suggested avoiding the Android warning displayed before redirection to a compatible browser; although this situation affected both groups, no participants in the passkey group reported this suggestion.

\begin{table*}[htbp]
	\centering
	\setlength{\tabcolsep}{5pt}
	\renewcommand{\arraystretch}{1.15}
	\resizebox{0.8\linewidth}{!}{%
		\begin{tabular}{lcccc}
			\hline
			\textbf{Code} &
			\textbf{Windows \% ($G1_{\text{passkey}}$)} &
			\textbf{Windows \% ($G2_{\text{password}}$)} &
			\textbf{Android \% ($G1_{\text{passkey}}$)} &
			\textbf{Android \% ($G2_{\text{password}}$)} \\
			\hline
			None                         										& 44.0 & 44.0 & 56.0 & 44.0 \\
			Improve the successful connection confirmation page  				& 8.0  & 24.0 & 12.0 & 0.0  \\
			Clarify difference between email and username						& 8.0  & 12.0 & 8.0  & 4.0  \\
			Minor UI suggestions (text size, design, etc.)						& 12.0 & 4.0  & 4.0  & 4.0  \\
			Provide more instructions/guidance during the process				& 12.0 & 4.0  & 4.0  & 8.0  \\
			Improve the speed of the connection									& 4.0  & 8.0  & 0.0  & 4.0  \\
			The UI should use university branding for trust 					& 4.0  & 0.0  & 0.0  & 4.0  \\
			Minor input fields suggestions										& 4.0  & 0.0  & 8.0  & 0.0  \\
			Integrate authentication with OS, avoid captive portal				& 0.0  & 4.0  & 8.0  & 4.0  \\
			Fix Android warning when redirecting to web browser					& 0.0  & 0.0  & 4.0  & 12.0 \\
			Use second-factor authentication									& 0.0  & 0.0  & 0.0  & 4.0  \\
			\hline
		\end{tabular}%
	}
	\caption{Suggestions by platform and group to the authentication task.}
	\label{tab:results:suggestions-login}
\end{table*}

\section{Discussion}
\label{sec:discussion}

This split-plot experiment compared the usability of passwords and passkeys in the context of a Wi-Fi captive portal. Although we observed a tendency for passkeys to be more usable than passwords, the differences were not statistically significant. Nevertheless, we found meaningful usability differences across authentication methods and operating systems, and we identified several issues participants encountered when connecting through the captive portal. In this section, we discuss the results and propose improvements to the user experience of captive-portal authentication.

\subsection{Passkeys vs. passwords in captive portals}


Effectiveness metrics were similar for both authentication methods, although they differed by operating system. Across both Windows and Android, task completion rates were comparable between $G1_{\text{passkey}}$ and $G2_{\text{password}}$, with Android showing lower completion overall. Users often abandoned the task due to errors on Android, with a higher abandonment rate in $G1_{\text{passkey}}$; on Windows, both methods achieved similar completion rates. We also observed that some first-attempt failures became successes in the retention test, suggesting that participants (especially those using passkeys on Android) learned from the first experience and performed better on the second attempt.

Efficiency measures showed a retention effect for both methods and platforms, as mean authentication times were lower in the second attempt. Considering that passkeys generally showed lower median authentication times (and reduced dispersion on the second attempt), we observe a tendency toward higher efficiency for passkeys compared to passwords, particularly on Android. Despite outliers caused by persistent errors, passkeys showed consistently lower median values and a greater reduction in dispersion after the first attempt. Participants in $G1_{\text{passkey}}$ also exhibited the most pronounced learning effect, while participants in $G2_{\text{password}}$ showed smaller improvements; notably, some participants who succeeded with passwords on the first attempt failed the second attempt.

We observed that participants in both groups faced errors in nearly 6 out of 10 tasks, mostly when using Android. Most errors were common for both groups. For instance, the most common was caused by the captive portal system not granting Internet access after successful authentication in a reasonable time. Other common errors were caused by the User Experience (UX) of the OS or the captive portal User Interface (UI), like the misleading ``Manage Account'' button.

However, some errors were specific to each method. Participants in $G1_{\text{passkey}}$ were affected by an error that aborted passkey registration or authentication when the action button was double-clicked. This issue can be mitigated by disabling the button after the first click (or debouncing the handler) until the ceremony completes. In contrast, the most common issues faced by participants in $G2_{\text{password}}$ involved username/password entry, including mistyping or forgetting credentials.

User satisfaction, measured using SUS at the end of the experiment and SEQ after each task, suggests broadly similar perceptions across groups. In Windows, about half of the participants in $G1_{\text{passkey}}$ assigned the highest SEQ score, compared with about one third of participants in $G2_{\text{password}}$. On Android, password authentication produced a larger proportion of low SEQ scores. The additional post-task questions also showed slightly higher mean ratings in $G1_{\text{passkey}}$, indicating that participants perceived passkey-based authentication as more intuitive on average than password-based authentication.

Qualitative results also suggest that participants using passkeys had a more positive overall experience. Reported advantages were diverse but largely consistent across groups: participants using passwords tended to mention generic benefits (e.g. familiarity), whereas participants using passkeys highlighted method-specific benefits such as ease of configuration and the ability to leverage biometric authentication. Regarding disadvantages, both groups agreed that the main issue was the friction associated with repeated authentication.

Overall, participants in both groups experienced usability problems related to the captive portal itself. We believe that these system-level issues reduced the observable effect size between authentication methods. \textbf{\textit{Answer to RQ1: Although we observe a tendency toward higher usability for passkeys in a captive portal, system-level errors (particularly those affecting connectivity after authentication) may have a larger impact on usability than the chosen authentication method.}}

\subsection{Usability of passkeys in Windows and Android}

The split-plot design allowed us to gather data for passkeys and passwords on both Windows and Android. The behaviour of the captive portal differs across operating systems, and the user experience of passkeys is additionally shaped by OS-specific UI and interaction patterns.

A key UX difference on Android is caused by its Captive Portal Detection (CPD) mechanism. After connecting to the network, Android typically opens a dedicated captive-portal mini-browser. However, this embedded browser is not compatible with the WebAuthn API and therefore cannot support passkey-based authentication. To mitigate this limitation, our captive-portal implementation detects the lack of WebAuthn support and redirects the user to a compatible browser (Chrome). This redirection requires leaving the CPD mini-browser and, as a security measure, Android blocks the transition until the user explicitly confirms it in an OS-level warning dialog.

We observed that this warning dialog often caused hesitation and confusion, with some participants interpreting it as a potential security risk. Moreover, unlike Windows (which usually opens the portal directly), Android relies on a notification-based CPD entry point, adding an additional interaction step. Participants frequently commented on this difference, and we believe it contributed to the lower perceived usability and higher failure rates observed on Android.

This is reflected in several usability metrics, such as completion rates, which differed substantially between Windows and Android. Specifically, Android presented the highest abandonment rate before completing authentication, likely due to the CPD flow and the browser redirection step. On Android, the learning effect had a notable impact, yielding greater effectiveness in the second attempt, especially for passkeys.

Consequently, Android also exhibited a higher overall error rate than Windows. The most frequent platform-specific error on Android was the OS notification stating ``Connected'' after redirection to the default browser, which occurred with a 32\% error rate. When asked about their general experience, a substantial number of participants mentioned Android-related issues driven by the CPD UX.

Regarding registration of the assigned authentication method, participants in both groups achieved a 100\% completion rate. Mean registration times were lower on Android than on Windows for both methods. We argue that the \textbf{\textit{answer to RQ2 is that passkeys were easy to configure on both platforms and achieved complete registration success, despite some errors caused by double-clicking the initiation button and suggestions for improved guidance (e.g. where the passkey is stored) and fallback methods such as a password.}}

\subsection{Roadmap for addressing usability issues in captive portals}


In this paper, we analysed the usability of a captive portal using two authentication methods. In this section, \textbf{\textit{we discuss the major usability issues and propose corresponding improvements to address RQ3}}.

\paragraph{\small{SLOW AND ERROR-PRONE CONNECTION}}
As reported by participants, the process of connecting to the Internet after authenticating through a captive portal is slow, introduces unfamiliar steps, and is prone to errors. We believe that authentication times can be improved by reducing the number of steps, minimising user-input friction, and improving the Captive Portal Detection (CPD) flow.

Regarding authentication, FIDO2 client-side (resident) credentials enable a usernameless experience in which users can authenticate by pressing a single button and completing platform biometric authentication. This flow eliminates the need to prompt users for a username and can trigger the operating system's FIDO2 dialog as the first authentication step, minimising user interaction. However, Android adds friction through its notification-based CPD flow, and the CPD mini-browser's lack of WebAuthn compatibility reduces the usability of passkey authentication. To improve connection speed, the operating system should, where possible, automatically open the captive portal in a WebAuthn-compatible browser via the CPD system, as Windows does.

\paragraph{\small{ABANDONMENT AFTER AUTHENTICATION}}
We detected that a substantial number of participants abandoned the connection process after successful authentication, misled by notifications or premature success messages. Participants expected immediate Internet connectivity after authentication, but delays in granting access caused frustration and led some to abandon the process.

For access control via captive portals, we believe it is critical to verify connectivity before confirming a successful connection to the user (or redirecting them to browsing). Such checks could be implemented via client-side scripting while displaying meaningful progress feedback. Additionally, the captive portal enforcement device should apply access policies as promptly as possible to reduce delays and, therefore, reduce the risk of abandonment.

\paragraph{\small{CONFUSING OR MISLEADING CONNECTION}}
In general, participants expressed confusion during the connection process, including during registration. Although in this laboratory setup registration and authentication happened sequentially within the same session, participants still reported that the overall connection process felt unfamiliar. In a real deployment, where users may already have an account, the flow may be easier to adopt; however, the captive-portal interaction itself remains a source of uncertainty.

Some aspects can be improved by providing additional guidance. For example, during passkey registration, users could be informed about the upcoming OS prompt (e.g. where the passkey will be stored) before it appears. Some users may also benefit from short, contextual explanations of passkeys (e.g. security benefits and account recovery in case of device loss). Regarding the OS, the CPD flow should reduce friction by prompting users automatically and by more clearly describing connection status in network settings, especially on Android.

\paragraph{\small{RE-AUTHENTICATION FRICTION}}
Re-authentication friction is an important factor for user satisfaction and adoption. To control network access, the access point must ensure that the device remains authorised, which typically requires the captive portal to maintain and validate an authenticated session. In this laboratory study, participants authenticated twice on each operating system, which represents a higher frequency than would typically occur in practice. Nevertheless, some participants expressed concerns about repeated authentication, which may affect adoption.

For this reason, captive portals should support automatic re-authentication without user interaction (e.g. using time-limited tokens) or maintain active sessions for a reasonable period. Faster, lower-friction flows based on passkeys and biometrics may help support more frequent re-authentication while minimising usability impact. In any case, the effect of authentication frequency should be studied in longitudinal deployments with different authentication methods.

\paragraph{\small{USER INPUT ERRORS}}
As in any interactive system, we observed software bugs and unexpected user input that caused errors during registration and authentication. These errors directly impacted the user experience and, in some cases, task completion. With passkeys, we observed failures caused by initiating registration/authentication twice before completion. With passwords, we observed usability issues related to credential entry (e.g. requests for a show-password option) and confusion about the username field.

As mentioned earlier, FIDO2 usernameless flows minimise user input and delegate more interaction to the operating-system UI. However, it is essential to ensure robust passkey registration and authentication before deployment to prevent confusion and to realise the potential usability benefits.

\subsection{Threats to validity}
Similarly to other studies in a university setting, our participants were young adults (e.g.~\cite{lyastani_is_2020}). Most were students enrolled in a degree related to computer science, which may have affected the generalisability of our results to other populations. For instance, participants with more technical knowledge may have been more resilient to errors and more tolerant of friction during the connection process. Despite this, we measured a mean ATI score of 4.4, which indicates a slightly high level of technology affinity in our sample. Future work should include more diverse samples to assess the generalisability of our findings.

The experiment was conducted in a controlled laboratory setting, which may not reflect real-world usage. For instance, in order to repeat tasks to measure retention, the captive portal connection sessions were configured to be ephemeral, which may have affected perceived usability compared to a real deployment. Furthermore, each participant was required to register and then authenticate twice on two different devices within the same laboratory session. This increased authentication frequency relative to everyday use and may have contributed to misunderstandings about the registration process. In future work, we plan to design longitudinal studies in real-world settings to assess the usability of passkeys and passwords in captive portals over time, separating the registration and authentication ceremonies and avoiding artificially high authentication frequency.

In this experiment, we focused on platform authenticators, which may not be generalisable to other types of FIDO2 authenticators (e.g. hardware security keys), but represent the current trend in the market. The use of pre-configured devices with a PIN may have affected the perception of the system, as participants could associate passkeys with a predefined shared secret (the device PIN), shaping their expectations and the perceived usability and security. To complement our findings, future longitudinal studies should consider using devices owned by participants, which may provide a more realistic experience when using the device lock, and therefore provide a better assessment of the usability of passkeys in this context.


Finally, our findings regarding the usability of passkeys on Android were influenced by the captive-portal mini-browser's lack of WebAuthn API support, which required redirection to a compatible browser. Although our captive-portal implementation attempts to mitigate this issue by automating this redirection transparently for the user, the OS-level warning message displayed by Android caused confusion and hesitation among participants, which may have affected the usability of passkeys on this platform. This issue is particularly relevant given the high error rates observed on Android. Future developments should consider this use case and explore better integration of the WebAuthn API into captive portals.

\section{Related work}
\label{sec:related-work}

Lyastani et al.~\cite{lyastani_is_2020} conducted the first laboratory study to compare FIDO2 and password authentication in terms of usability. Their between-group experiment used FIDO2 hardware security keys for passwordless authentication. They found that passwordless authentication with security keys was both more usable and more acceptable than passwords. Building on this work, we conduct a comparative usability study of FIDO2 authentication and passwords. However, our experiment has two main differences.

First, our study involves a different scenario: authentication in a captive portal when users connect to a Wi-Fi network. In~\cite{rivera-dourado_novel_2024} we introduced FIDO2CAP, a captive-portal authentication protocol based on FIDO2 that enables users to authenticate to a Wi-Fi network using passkeys. Building on this prior work, our experiment uses a split-plot design to study the usability of passkeys compared to passwords in a captive portal.

Second, our FIDO2 usability experiment does not use hardware security keys; instead, it focuses on platform authenticators. Oogami et al.~\cite{oogami_observation_2020} examined the usability potential of FIDO2 platform authenticators on Android smartphones. In their experiment, participants used their fingerprint to authenticate on a WebAuthn-compatible website using their own Android devices, and some found the user interface misleading. The first to compare the usability of platform-based authenticators with hardware security keys were Würschung et al.~\cite{wursching_fido2_2023}, who compared both procedures when using FIDO2 authentication on a smartphone. They found that users slightly preferred platform authenticators, particularly for frequently used online accounts. Accordingly, our scenario relies on platform-based passkeys: participants used the built-in authenticators on both Windows and Android to authenticate through the captive portal.

Both of these cited experiments were conducted in controlled laboratory sessions, whereas other FIDO2 usability studies have used longitudinal designs that include initial registration followed by daily authentication. Owens et al.~\cite{owens_user_2021,owens_framework_2020} studied the usability of Android devices as FIDO2 authenticators, comparing their FIDO2 prototype (Neo) with passwords in a between-group study. Neo involved cross-device authentication, sending a push notification to the user's smartphone to approve authentication on a laptop, which required an initial configuration phase. In this two-week study, users authenticated daily to a web application on their laptop using their assigned method. They found that Neo yielded fewer authentication errors during daily use, but that initial setup shaped overall perception.

Similarly, Reynolds et al.~\cite{reynolds_tale_2018} measured the usability of an external authenticator by using a Yubico security key as a second factor with Google and Facebook, based on the FIDO Universal Second Factor (U2F) standard~\cite{ciolino_two_2019}, a predecessor of FIDO2. They found that setup was perceived negatively, whereas daily use was pleasant for most users. For this reason, our experiment does not involve external authenticators that could affect the configuration experience; instead, we use a registration flow designed to be as simple as possible with platform authenticators in a controlled environment.

When using FIDO2 authenticators as a first factor, deployments typically require User Verification (UV), which involves local authentication to approve the operation. Farke et al.~\cite{farke_you_2020} conducted a qualitative study on participants' perceptions of a YubiKey security key protected with a PIN as a first-factor authentication method. In an enterprise setting~\cite{kepkowski_challenges_2023}, employees used the hardware key for four weeks while keeping an authentication diary and were interviewed at the end of the experiment. Some users did not perceive a clear advantage over their prior password routines, and the PIN requirement introduced additional steps.

In contrast, Lassak et al.~\cite{lassak_totps_2025} reported different outcomes in a corporate setting. They compared the usability of the YubiKey Bio security key with an existing password+TOTP scheme, evaluating the two UV methods supported by the YubiKey Bio: fingerprint and PIN. When the security key was protected by fingerprint, authentication was faster than the password+2FA baseline. However, PIN-based use was not notably faster. Overall, they found that fingerprint-protected FIDO2 security keys significantly improved user satisfaction compared to password+2FA.

In our study, the Windows laptop and the Android smartphone are PIN-protected. During FIDO2 authentication using the platform authenticator, users must enter this PIN to complete UV. Unlike the cited enterprise studies, our work does not rely on an external security key; instead, we use built-in platform authenticators. To reduce cognitive load, participants unlocked the device with the four-digit PIN before starting the experiment tasks.

Finally, several studies have explored the usability of access-control systems in Wi-Fi networks. Brown et al.~\cite{brown_multinet_2013} presented a system for connecting devices to domestic networks using QR codes. To assess usability, they conducted a within-subject study comparing their system with Wi-Fi Protected Setup (WPS). More recently, Hasan et al.~\cite{hasan_seqr_2025} introduced a QR-code-based system to ease the configuration of corporate environments protected by Extensible Authentication Protocol (EAP) methods. They compared two standard manual configuration approaches with their SeQR system in a within-subject experiment.

Captive portal systems have also been evaluated in usability experiments. Budde et al.~\cite{budde_comparative_2014} evaluated the usability of six innovative captive portal systems, seeking alternatives to passwords for Wi-Fi access. While our goal is similarly to explore replacements for passwords, we evaluate FIDO2 as an alternative: a standard authentication technology widely adopted in web authentication and backed by the FIDO Alliance.

\section{Conclusion}
\label{sec:conclusion}

In this paper, we studied the usability of passkeys and passwords as authentication methods in a captive portal. We designed a split-plot experiment with 50 participants who used both Android and Windows platforms, which influenced their overall user experience. To conduct the experiment, we configured a Wi-Fi laboratory setup with a laptop and a smartphone, together with an OpenWRT wireless router implementing the FIDO2CAP captive portal authentication protocol~\cite{rivera-dourado_novel_2024}.

We found a tendency for passkeys to be more usable than passwords when connecting through a captive portal (RQ1). However, the differences were not statistically significant. Participants experienced a high error rate caused by limitations of captive portals, which affected user experience regardless of authentication method. In particular, connectivity problems after successful authentication caused a substantial number of errors as well as outliers in time-on-task and task abandonment in both groups.

Passkeys were easy to configure on both Windows and Android, achieving complete registration success (RQ2). We identified a software bug whereby double-pressing the initiation button could abort passkey registration. We also observed usability issues specific to passkey authentication on Android: the CPD mini-browser is not compatible with the WebAuthn API required for passkeys, which forced the system to redirect users to a compatible browser and negatively affected user experience.

Based on this experiment, we propose a set of recommendations to improve captive-portal usability. These include adopting passkey flows that support usernameless authentication, improving CPD behaviour, refining user-interface feedback, and studying how different re-authentication frequencies affect system usability.

Finally, future work should validate these improvements in follow-up experiments, including re-authentication frequency, FIDO2 usernameless flows, and enhancements to CPD systems. We believe that longitudinal studies in which participants connect to everyday Wi-Fi networks using already registered accounts will provide additional insights into the usability of authentication methods in captive portals.

Such studies can help improve the usability of passkeys, support adoption in new use cases such as network authentication, and contribute to the transition toward more secure and ubiquitous passwordless authentication systems.

{\small \section*{CRediT authorship contribution statement}
\textbf{Martiño Rivera-Dourado}: Conceptualization, Methodology, Software, Investigation, Data Curation, Writing – original draft, Writing review \& editing, Visualization, Project administration. \textbf{Rubén Pérez-Jove}: Investigation, Data Curation, Writing – review \& editing, Visualization. \textbf{Alejandro Pazos}: Resources, Supervision, Project administration, Funding acquisition. \textbf{Jose Vázquez-Naya}: Methodology, Resources, Writing – review \& editing, Supervision, Project administration.}

{\small \section*{Declaration of competing interest}
The authors declare that they have no known competing financial interests or personal relationships that could have appeared to influence the work reported in this paper.}

{\small \section*{Acknowledgements}

This work was founded by EU and ``Xunta de Galicia'' (Spain), grant ED431C 2022/46–Competitive Reference Groups GRC. This work was also supported by CITIC, as a center accredited for excellence within the Galician University System and a member of the CIGUS Network, which receives subsidies from the Department of Education, Science, Universities, and Vocational Training of the ``Xunta de Galicia''. Additionally, CITIC is co-financed by the EU through the FEDER Galicia 2021-27 operational program (Ref. ED431G 2023/01). The work is also founded by the ``Formación de Profesorado Universitario'' (FPU) grant from the Spanish Ministry of Universities to Martiño Rivera Dourado (Grant FPU21/04519) and to Rubén Pérez Jove (Grant FPU22/04418). Funding for open access charge: Universidade da Coruña/CISUG.}

{\small \section*{Data availability}

Data will be made available on request.

{\small \section*{Declaration of generative AI and AI-assisted technologies in the manuscript preparation process}

During the preparation of this work the author(s) used generative LLMs in order to proofread and to improve readability and presentation of the manuscript. After using this tool/service, the author(s) reviewed and edited the content as needed and take(s) full responsibility for the content of the published article.

\printbibliography

\clearpage
\appendix

\section{Pre-experiment Survey}
\label{app:pre-experiment-survey}

This appendix presents the questionnaire administered to participants before the experiment.

\small{\textit{Note: The original questionnaires were administered in Spanish. The text below represents the English translation of the instruments used during the experiment.}}

\small{

\subsection*{Demographics}

\begin{itemize}
	\item \textbf{Select your age range}
	\begin{itemize}
		\item 18--25 years old
		\item 26--35 years old
		\item 35--45 years old
		\item 45--55 years old
		\item 55--65 years old
		\item More than 65 years old
	\end{itemize}
	
	\item \textbf{Select your gender}
	\begin{itemize}
		\item Masculine
		\item Feminine
		\item Non-binary / Other
		\item Prefer not to say
	\end{itemize}
	
	\item \textbf{Please, select your education level}
	\begin{itemize}
		\item Primary education
		\item Secondary education
		\item Upper secondary education / Basic professional training
		\item Higher professional training
		\item Bachelor’s degree / Undergraduate degree
		\item Master’s degree
		\item Doctoral degree (PhD)
\end{itemize}

	\item \textbf{Indicate your field of work or study}
	\begin{itemize}
		\item Technological background: IT, software development, data analysis and/or telecommunications
		\item Other engineering fields
		\item Other scientific fields
		\item Law, management, finance or other related fields
		\item Arts, history, sociology, education or other related fields
		\item Other (please specify)
	\end{itemize}
\end{itemize}

\subsection*{Affinity for Technology Interaction (ATI)}

For each of the following statements, participants indicated their level of agreement
using a 6-point Likert scale, where \textbf{1 = Completely disagree} and
\textbf{6 = Completely agree}.

\medskip

\noindent\textbf{01.} I like to occupy myself in greater detail with technical systems.

{\centering\small
	1~$\bigcirc$\quad 2~$\bigcirc$\quad 3~$\bigcirc$\quad
	4~$\bigcirc$\quad 5~$\bigcirc$\quad 6~$\bigcirc$\par}

\medskip
\noindent\textbf{02.} I like testing the functions of new technical systems.

{\centering\small
1~$\bigcirc$\quad 2~$\bigcirc$\quad 3~$\bigcirc$\quad
4~$\bigcirc$\quad 5~$\bigcirc$\quad 6~$\bigcirc$\par}

\medskip
\noindent\textbf{03.} I predominantly deal with technical systems because I have to.

{\centering\small
1~$\bigcirc$\quad 2~$\bigcirc$\quad 3~$\bigcirc$\quad
4~$\bigcirc$\quad 5~$\bigcirc$\quad 6~$\bigcirc$\par}

\medskip
\noindent\textbf{04.} When I have a new technical system in front of me, I try it out intensively.

{\centering\small
1~$\bigcirc$\quad 2~$\bigcirc$\quad 3~$\bigcirc$\quad
4~$\bigcirc$\quad 5~$\bigcirc$\quad 6~$\bigcirc$\par}

\medskip
\noindent\textbf{05.} I enjoy spending time becoming acquainted with a new technical system.

{\centering\small
1~$\bigcirc$\quad 2~$\bigcirc$\quad 3~$\bigcirc$\quad
4~$\bigcirc$\quad 5~$\bigcirc$\quad 6~$\bigcirc$\par}

\medskip
\noindent\textbf{06.} It is enough for me that a technical system works; I don’t care how or why.

{\centering\small
1~$\bigcirc$\quad 2~$\bigcirc$\quad 3~$\bigcirc$\quad
4~$\bigcirc$\quad 5~$\bigcirc$\quad 6~$\bigcirc$\par}

\medskip
\noindent\textbf{07.} I try to understand how a technical system exactly works.

{\centering\small
1~$\bigcirc$\quad 2~$\bigcirc$\quad 3~$\bigcirc$\quad
4~$\bigcirc$\quad 5~$\bigcirc$\quad 6~$\bigcirc$\par}

\medskip
\noindent\textbf{08.} It is enough for me to know the basic functions of a technical system.

{\centering\small
1~$\bigcirc$\quad 2~$\bigcirc$\quad 3~$\bigcirc$\quad
4~$\bigcirc$\quad 5~$\bigcirc$\quad 6~$\bigcirc$\par}

\medskip
\noindent\textbf{09.} I try to make full use of the capabilities of a technical system.

{\centering\small
1~$\bigcirc$\quad 2~$\bigcirc$\quad 3~$\bigcirc$\quad
4~$\bigcirc$\quad 5~$\bigcirc$\quad 6~$\bigcirc$\par}

\subsection*{Device Usage}

\begin{itemize}
	\item \textbf{Do you use a Windows computer or laptop frequently?}
	\begin{itemize}
		\item No, I do not use a computer or laptop
		\item No, I do not use Windows frequently
		\item Yes, but once a week
		\item Yes, 2--4 times a week
		\item Yes, almost every day
	\end{itemize}
	
	\item \textbf{Do you use an Android smartphone frequently?}
	\begin{itemize}
		\item No, I do not have a smartphone
		\item No, I have an iPhone
		\item Yes, but once a week
		\item Yes, 2--4 times a week
		\item Yes, almost every day
	\end{itemize}
\end{itemize}

\subsection*{Authentication Knowledge and Practices}

\begin{itemize}
	\item \textbf{How familiar are you with two-step authentication (2FA)?}
	\begin{itemize}
		\item I do not know what it is
		\item It is familiar to me, but I am not sure
		\item Yes, I know what it is, but I have never used it
		\item Yes, I know it and I use it
	\end{itemize}
	
	\item \textbf{Which authentication methods do you use in your online accounts?} (Select all that apply)
	\begin{itemize}
		\item Passwords
		\item SMS code
		\item Code sent to email
		\item Authenticator app
		\item Passkeys or physical security keys/tokens
		\item Numeric PIN
		\item Digital certificates
		\item DNIe or electronic ID
		\item Other (please indicate)
	\end{itemize}
	
	\item \textbf{How do you lock your laptop or computer?}
	\begin{itemize}
		\item Password
		\item Numeric PIN
		\item Fingerprint
		\item Face recognition
		\item Security key or token
		\item The device logs in automatically when booting
	\end{itemize}
	
	\item \textbf{How do you lock your smartphone?}
	\begin{itemize}
		\item Password
		\item Numeric PIN
		\item Fingerprint
		\item Face recognition
		\item I do not lock my smartphone (slide to unlock)
	\end{itemize}
\end{itemize}

}

\section{Post-task survey}
\label{app:post-task-survey}

After completing the authentication task on both Windows and Android devices, participants answered the following post-task survey.

\small{

\subsection*{Perception questionnaire}

\medskip
\noindent\textbf{01.} The Wi-Fi login process was direct and intuitive.

{\centering\small
	{\scriptsize Strongly disagree --}\quad
	1~$\bigcirc$\quad 2~$\bigcirc$\quad 3~$\bigcirc$\quad
	4~$\bigcirc$\quad 5~$\bigcirc$\quad
	{\scriptsize -- Strongly agree}\par}

\medskip
\noindent\textbf{02.} I felt confident completing the Wi-Fi login.

{\centering\small
	{\scriptsize Strongly disagree --}\quad
	1~$\bigcirc$\quad 2~$\bigcirc$\quad 3~$\bigcirc$\quad
	4~$\bigcirc$\quad 5~$\bigcirc$\quad
	{\scriptsize -- Strongly agree}\par}

\medskip
\noindent\textbf{03.} I felt that the Wi-Fi login method was secure and trustworthy.

{\centering\small
	{\scriptsize Strongly disagree --}\quad
	1~$\bigcirc$\quad 2~$\bigcirc$\quad 3~$\bigcirc$\quad
	4~$\bigcirc$\quad 5~$\bigcirc$\quad
	{\scriptsize -- Strongly agree}\par}

\medskip
\noindent\textbf{04.} The amount of effort required to complete the Wi-Fi login was reasonable.

{\centering\small
	{\scriptsize Strongly disagree --}\quad
	1~$\bigcirc$\quad 2~$\bigcirc$\quad 3~$\bigcirc$\quad
	4~$\bigcirc$\quad 5~$\bigcirc$\quad
	{\scriptsize -- Strongly agree}\par}

\medskip
\noindent\textbf{05.} Overall, how easy or difficult was it to complete the Wi-Fi login?

{\centering\small
	{\scriptsize Strongly disagree --}\quad
	1~$\bigcirc$\quad 2~$\bigcirc$\quad 3~$\bigcirc$\quad
	4~$\bigcirc$\quad 5~$\bigcirc$\quad
	{\scriptsize -- Strongly agree}\par}

\subsection*{SEQ}

\begin{enumerate}
	\item Overall, how easy or difficult was it to complete the Wi-Fi login?
\end{enumerate}

{\centering\small
	{\scriptsize Very difficult --}\quad
	1~$\bigcirc$\quad 2~$\bigcirc$\quad 3~$\bigcirc$\quad
	4~$\bigcirc$\quad 5~$\bigcirc$\quad 6~$\bigcirc$\quad
	7~$\bigcirc$\quad
	{\scriptsize -- Very easy}\par}

\subsection*{Open-ended questions}

\begin{enumerate}
	\item Do you have any suggestion to improve the login process?
	\item Do you have any suggestion to improve the registration or configuration process?
\end{enumerate}

}

\newpage

\section{Post-experiment Survey}
\label{app:post-experiment-survey}

This questionnaire was administered after participants completed the authentication tasks.

\subsection*{System Usability Scale (SUS)}

Participants rated the following statements using a 5-point Likert scale
\textbf{(1 = Strongly disagree, 5 = Strongly agree)}.

\medskip
\noindent\textbf{01.} I think that I would like to use this system frequently.

\medskip
\noindent\textbf{02.} I found the system unnecessarily complex.

\medskip
\noindent\textbf{03.} I thought the system was easy to use.

\medskip
\noindent\textbf{04.} I think that I would need the support of a technical person to be able to use this system.

\medskip
\noindent\textbf{05.} I found the various functions in this system were well integrated.

\medskip
\noindent\textbf{06.} I thought there was too much inconsistency in this system.

\medskip
\noindent\textbf{07.} I would imagine that most people would learn to use this system very quickly.

\medskip
\noindent\textbf{08.} I found the system very cumbersome to use.

\medskip
\noindent\textbf{09.} I felt very confident using the system.

\medskip
\noindent\textbf{10.} I needed to learn a lot of things before I could get going with this system.

\subsection*{Open-ended Questions}

\begin{itemize}
	\item How would you describe your general experience with the authentication method you used?
	\item What advantages do you see with using this authentication method?
	\item What disadvantages do you see with using this authentication method?
\end{itemize}

\newpage
\clearpage

\section{Observed errors}
\label{appendix:observed-errors}

As explained in Section~\ref{sec:quantitative-results}, some of the errors observed during the experiment were caused by User Interface (UI) elements that influenced user behaviour. Figure~\ref{fig:ui-elements-observed-errors} includes screenshots of these UI elements.

\begin{enumerate}[label=(\alph*)]
	\item \textbf{Android warning:} Within the Captive Portal Detection (CPD) mini-browser, Android displays a warning message when the captive portal attempts to open a browser to complete authentication. The translated message claims that \textit{``The network you are trying to connect to requests to open another app. For example, it is possible that the login requires a specific app for authentication. Continue anyway with the browser''}. This message caused confusion and hesitation among participants, which may have contributed to the high error rates observed on Android.
	\item \textbf{Android ``Connected'' notification:} After the user confirms opening the external browser, Android displays a \textit{``Connected''} notification. Normally, the user should press \textit{``Login''} on the page titled \textit{``Hi! Ready to login?''}. However, the Android pop-up notification misled some participants into thinking they had already completed the connection process, leading to task abandonment or frustration when checking Internet connectivity.
	\item \textbf{Android ``Manage account'' button:} After successful authentication, the captive portal redirects users to a confirmation page with the message \textit{``Great! You are now connected to the Internet''} that includes the \textit{``Logout''} and the \textit{``Manage account''} buttons. Some participants clicked the \textit{``Manage account''} button, which caused an error because it was not part of the intended flow.
\end{enumerate}

\begin{figure}[ht!]
	\centering
	\begin{subfigure}[b]{0.30\linewidth}
		\centering
		\includegraphics[width=\linewidth]{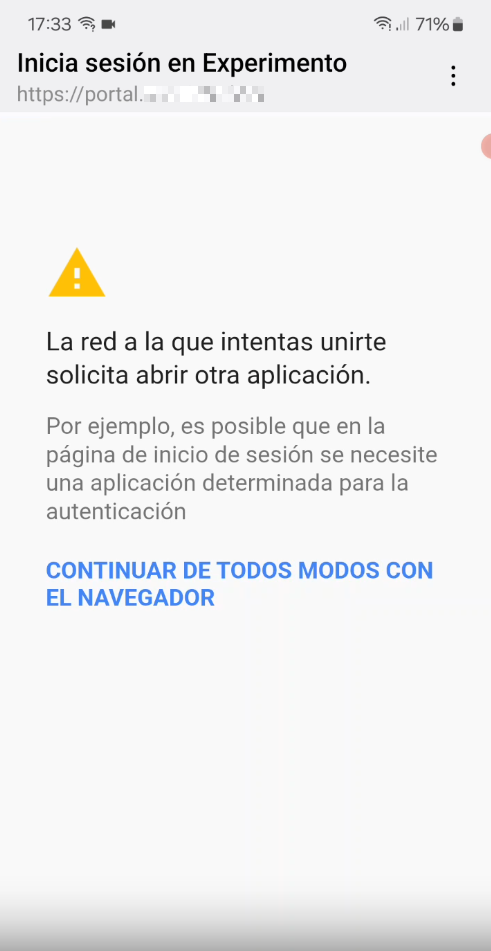}
		\caption{Android warning before opening browser, with a continue call to action.}
	\end{subfigure}
	\hfill
	\begin{subfigure}[b]{0.30\linewidth}
		\centering
		\includegraphics[width=\linewidth]{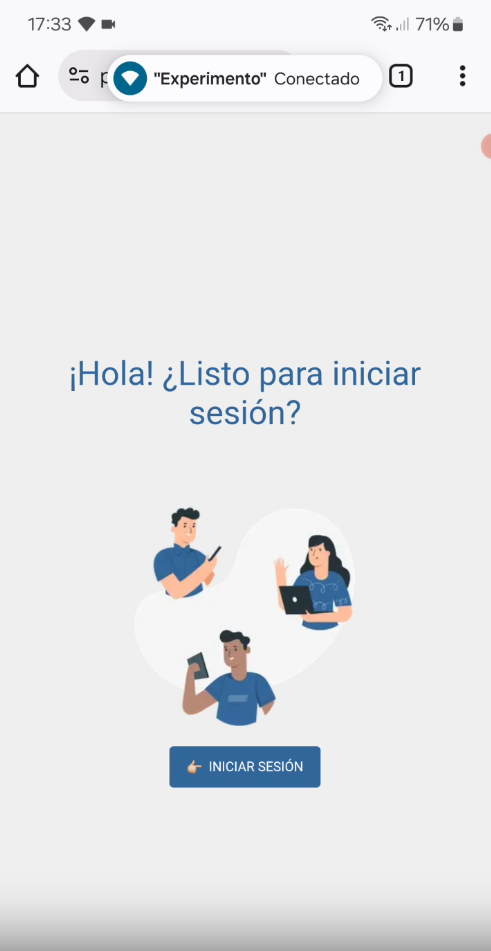}
		\caption{\textit{``Connected''} notification on Android before authentication.}
	\end{subfigure}
	\hfill
	\begin{subfigure}[b]{0.30\linewidth}
		\centering
		\includegraphics[width=\linewidth]{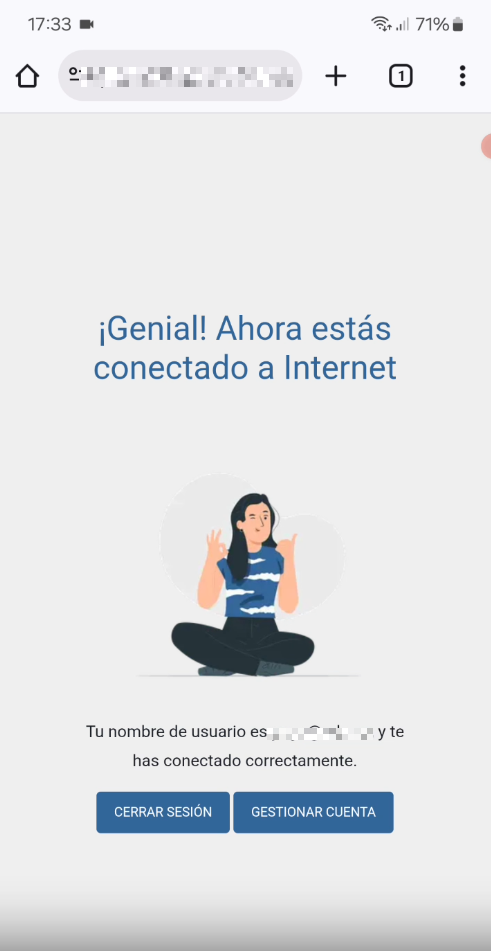}
		\caption{\textit{``Manage account''} button on the confirmation page after authentication.}
	\end{subfigure}
	\caption{UI elements causing some of the observed errors. Screenshots from a recorded session of a participant on Android, originally in Spanish. Translated UI text is included in the appendix. }
	\label{fig:ui-elements-observed-errors}
\end{figure}

\onecolumn
\section{Codebooks}
\label{appendix:codebooks}

\subsection{Post-experiment codebooks}
\label{appendix:codebooks:post-experiment}

\begin{table*}[!ht]
	\centering
	\caption{Codebook for post-experiment general user experience questionnaire.}
	\vspace{8pt}
	\resizebox{0.8\textwidth}{!}{%
		\begin{tabular}{l@{\hspace{10pt}}c@{\hspace{15pt}}p{6cm}@{\hspace{15pt}}p{6cm}}
			\toprule
			\textbf{Code} & \textbf{Freq.} & \textbf{Description} & \textbf{Example} \\
			\midrule
			Ease of use 			& 25 & Positive comments regarding ease of use and smooth interaction. & \textit{``The overall experience is good; it seems comfortable and intuitive to use [...]''} \\ \midrule
			General positive 		& 19 & General positive evaluation of the overall experience. & \textit{``A good experience, [...]''} \\ \midrule
			Android issues 			& 12 & Issues specifically attributed to the Android platform. & \textit{``although it's true that on Android it's more tedious than on Windows and it took some effort to get it to connect the first time [...]''} \\ \midrule
			Confusion	 			& 10 & Confusion or lack of clarity during interaction with the system. & \textit{``[...] I feel that when it doesn't connect on the first try, [...] it could confuse users with less experience or technical knowledge''} \\ \midrule
			Not reliable	 		& 7 & Perceived technical issues, errors, or unreliable behaviour. & \textit{``Quite cumbersome and unreliable when it comes to obtaining an internet connection.''} \\ \midrule
			Secure 					& 6 & Positive perception of security or trustworthiness. & \textit{``Easy to use and more secure than a normal Wi-Fi password.''} \\ \midrule
			Familiar or intuitive	& 4 & The system feels familiar or intuitive due to prior experience. & \textit{``[...] a fast, fairly intuitive, and overall simple approach that anyone who has logged into an application with a username and password will be able to use quite easily.''} \\ \midrule
			Excessive effort 		& 4 & Perception that the system requires excessive effort. & \textit{``Very little user feedback; very confusing and cumbersome [...]''} \\ \midrule
			Fast	 				& 3 & Perception that the system is fast or time-efficient. & \textit{``[...] it's very convenient and fast in both cases [Android and Windows].''} \\ \midrule
			Friction auth. repetition & 2 & Frustration due to repeated authentication requests. & \textit{``[...] having to log in every time you want to connect to the Wi-Fi seems very cumbersome if it's a network you connect to regularly''} \\ \midrule
			Technical barrier 		& 2 & The system is perceived as too technical or complex. & \textit{``Slightly technical because it's not the usual way to connect to the internet [...]''} \\ \midrule
			Security concerns 		& 2 & Security concerns or distrust related to data, biometrics, or third parties. & \textit{``[...] although it doesn't inspire confidence in me that it uses the computer's security system [...]''} \\ \midrule
			Windows issues 			& 1 & Issues specifically attributed to the Windows platform. & \textit{``[...] I'm also wary that [...] on Windows it connects to the Internet before accessing the registration website.''} \\ \midrule
			Passkey issues 			& 1 & Problems related to passkey verification or usage. & \textit{``[...] at first it said there was a passkey verification problem [...]''} \\ \midrule
			General negative		& 1 & General negative assessment of the experience. & \textit{[...] a disaster.} \\ \midrule
			Confusion connection order & 1 & Confusion caused by Wi-Fi connection occurring before authentication. & \textit{``[...] it connected to the Wi-Fi even before entering the username or entering the phone's PIN.''} \\ \midrule
			Learning effect 		& 1 & Improved experience over time due to familiarity. & \textit{``Good, especially the second time, when you can avoid the mistakes you made the first time''} \\ \midrule
			None 					& 1 & No specific comments or neutral feedback. & \textit{``Different.''} \\
			\bottomrule
	\end{tabular}}
	
\end{table*}

\begin{table*}[!ht]
	\centering
	\caption{Codebook for perceived advantages of the authentication system.}
	\vspace{8pt}
	\resizebox{0.8\textwidth}{!}{%
		\begin{tabular}{l@{\hspace{10pt}}c@{\hspace{10pt}}p{6cm}@{\hspace{15pt}}p{6cm}}
			\toprule
			\textbf{Code} & \textbf{Freq.} & \textbf{Description} & \textbf{Example} \\
			\midrule
			Secure 						& 13 & General perception of increased security. & \textit{``More security''} \\ \midrule
			Easy or simple to use 		& 13 & Authentication process is easy and intuitive. & \textit{``Very fast and simple, anyone can use it without needing to know about computers.''} \\ \midrule
			Secure access control 		& 11 & Improved control over who can access the network. & \textit{``[...] in organisations, where it helps limit internet access and provides control over the network.''} \\ \midrule
			Fast 						& 9 & Faster authentication compared to traditional methods. & \textit{``It's faster than the one currently used at the university and doesn't require installing anything.''} \\ \midrule
			Perceived security 			& 6 & Strong subjective feeling of security and trust. & \textit{``It gives a greater sense of security''} \\ \midrule 
			Comfortable 				& 5 & Comfortable and smooth user experience. & \textit{``It's more convenient and secure than using passwords [...]''} \\ \midrule
			Familiarity 				& 3 & Uses authentication methods familiar to most users. & \textit{``[...] The Wi-Fi authentication in this experiment is quite similar to most other applications and websites that require a registered account [...]''} \\ \midrule
			Avoids pre-shared credentials & 3 & Avoids sharing a common pre-shared key. & \textit{``To be able to have a unique password per user for each Wi-Fi network, instead of a single one for everyone.''} \\ \midrule
			No manual configuration 	& 3 & No need for manual configuration or certificate installation. & \textit{``Simplicity compared to having to manually install certificates''} \\ \midrule
			Can benefit from biometrics	& 2 & Increased security due to biometric authentication mechanisms. & \textit{``4-digit PINs and fingerprints (which I understand would be compatible with this system) are much faster than entering passwords. [...] using biometric authentication methods [...] may be more secure [...]''} \\ \midrule
			No need to remember password & 2 & Eliminates the need to remember or type passwords. & \textit{``An alternative for a network you use frequently without having to remember the entire password.''} \\ \midrule
			UI attractive 				& 2 & Modern, clean, and visually appealing interface. & \textit{``[...] with a very user-friendly and attractive interface, displaying information and buttons clearly.''} \\ \midrule
			Useful for enterprise 		& 2 & Suitable for corporate or enterprise environments. & \textit{``[...] with great potential for corporate networks, where [...] corporate accounts are used.''} \\ \midrule
			Useful for events 			& 2 & Well-suited for temporary or public access contexts. & \textit{``[...] works well for an event where people have already registered, and then log in to connect, preventing outsiders from accessing the network [...]''} \\ \midrule
			None 						& 2 & No specific advantages identified. & \textit{``I'm not aware of any advantages it could provide me.''} \\
			\bottomrule
	\end{tabular}}
	
\end{table*}

\begin{table*}[!ht]
	\centering
	\caption{Codebook for perceived disadvantages of the authentication system.}
	\vspace{8pt}
	\resizebox{0.8\textwidth}{!}{%
		\begin{tabular}{l@{\hspace{10pt}}c@{\hspace{10pt}}p{6cm}@{\hspace{15pt}}p{6cm}}
			\toprule
			\textbf{Code} & \textbf{Freq.} & \textbf{Description} & \textbf{Example} \\
			\midrule
			Friction caused by repetition & 7 & Repetitive or frequent authentication causing user fatigue. & \textit{``Somewhat tedious if it has to be done daily; it would be convenient if it happened automatically, or by asking to restart the session every few months.''} \\ \midrule
			Authentication errors 		& 5 & Errors or failures during authentication. & \textit{``[...] with failures in the authentication system when connecting, which can make the user lose patience [...]''} \\ \midrule
			Security risk 				& 5 & Perceived risk of phishing, spoofing, or security attacks. & \textit{``Passwords could be leaked and someone could impersonate you.''} \\ \midrule
			Not familiar authentication & 4 & Authentication method is unfamiliar or novel to users. & \textit{``It's new, so you need to learn its particularities.''} \\ \midrule
			Access control risk 		& 4 & Fear of unauthorised access to the network. & \textit{``I'm not sure what the system would be to ensure that only authorised personnel access the Wi-Fi network [...]''} \\ \midrule
			Too many steps 				& 4 & The authentication process is perceived as too long or complex. & \textit{``More steps to connect, but the increase in security makes it worthwhile.''} \\ \midrule
			Technical barrier 			& 4 & The system is difficult for non-technical or older users. & \textit{``The average user would give up and get frustrated. For example, my mother would just ask me -- she's 50 years old.''} \\ \midrule
			Perceived insecurity 		& 3 & General feeling that the system is not secure enough. & \textit{``It might convey a sense of insecurity due to lacking a familiar authentication method like the traditional password.''} \\ \midrule
			Requires an account 		& 3 & Requirement to create an account limits accessibility. & \textit{``The need to create a user would limit network access for temporary visitors or require creating a user for single-use only.''} \\ \midrule
			Privacy concern 			& 2 & Concerns about personal data, tracking, or third-party access. & \textit{``You don’t see step by step what you’re doing or your data; it seems to me like you’re giving data and trust to an external party [...]''} \\ \midrule
			Insecure 1FA passkey 		& 2 & Perception that passkey-only authentication is insufficient. & \textit{``The lack of a password.''} \\ \midrule
			Slow or inefficient 		& 2 & Authentication process takes too long. & \textit{``Only the time -- it takes a bit longer to connect to a network or log in.''} \\ \midrule
			UI confusing 				& 2 & Confusing or unclear user interface elements. & \textit{``[...] on Android, a brief error screen appears during registration that shouldn't show, since the registration process was successful.''} \\ \midrule
			Requires specific hardware 	& 1 & Dependence on specific hardware (e.g. camera or biometric sensors). & \textit{``[...] some methods require special hardware to read fingerprints, do facial recognition, etc., which may not be available on some devices [...]''} \\ \midrule
			Insecure 1FA password 		& 1 & Perception that password-only authentication is insecure. & \textit{``Using only a password for authentication can be a bit insecure [...]''} \\ \midrule
			Remember password 			& 1 & Difficulty remembering or managing passwords. & \textit{``Since we need too many passwords for too many online accounts, we can sometimes forget some of them. [...]''} \\ \midrule
			Connection confusing 		& 1 & Uncertainty about connection status after authentication. & \textit{``especially if the connection process isn't clarified [...]''} \\ \midrule
			Friction to connect 		& 1 & Friction related to connecting to the Wi-Fi network itself. & \textit{[...] very inconvenient [...] to have to enter a username and password just to connect to a Wi-Fi network [...]} \\ \midrule
			None 						& 14 & No disadvantages perceived. & \textit{``Honestly, I don't think there is any disadvantage.''} \\
			\bottomrule
	\end{tabular}}
\end{table*}

\clearpage
\subsection{Post-task questionnaire codebooks}
\label{appendix:codebooks:post-task}

\begin{table*}[!ht]
	\centering
	\caption{Codebook for post-task login suggestions questionnaire.}
	\vspace{8pt}
	\resizebox{0.8\textwidth}{!}{%
		\begin{tabular}{l@{\hspace{10pt}}c@{\hspace{15pt}}p{6cm}@{\hspace{15pt}}p{6cm}}
			\toprule
			\textbf{Code} & \textbf{Freq.} & \textbf{Description} & \textbf{Example} \\
			\midrule
			Error handling guidance 		& 15 & Requests for clearer error messages and guidance when login failures occur. & \textit{``[...] and in case of an error or lack of connection, provide the user with more information about what’s happening.''} \\ \midrule
			Improve confirmation page 		& 11 & Need for clearer feedback confirming whether the connection or login was successful. & \textit{``Confirm more clearly that the process has been completed successfully and that you can start browsing.''} \\ \midrule
			Username / email confusion 		& 8 & Confusion between username and email as required login credentials. & \textit{``Clarify that the username is actually the email address.''} \\ \midrule
			More instructions / guidance 	& 7 & Requests for clearer instructions, explanations, or guidance during the login process. & \textit{``For example, that the web page provides information about the next steps, from the pop-up to the settings you need to configure.''} \\ \midrule
			UI visual improvements 			& 6 & Suggestions to improve visual design or usability of the login interface. & \textit{``Slightly larger text.''} \\ \midrule
			Improve speed 					& 4 & Perception that the login process is slow or could be faster. & \textit{``That it were faster. Sometimes it felt like it took a while to load, as if something had failed, but that wasn't the case.''} \\ \midrule
			Fix Android warning in redirection & 4 & Confusion or concern caused by Android security warnings or redirections. & \textit{``A message asking whether you want to trust a redirection to another browser should not appear. You should be able to do it in the default one.''} \\ \midrule
			Avoid using a captive portal 	& 4 & Desire to integrate login directly without relying on an external captive portal. & \textit{``The device could automatically connect to the Wi-Fi if it's an open network.''} \\ \midrule
			Confusion of connectivity 		& 4 & Uncertainty about whether the device is connected automatically or requires manual action. & \textit{``Yes, not entirely intuitive, since the first time the session didn't login and I had to do it again once I realised.''} \\ \midrule
			Confusion due to manage acc. button & 4 & Lack of awareness about account management or configuration options. & \textit{``In my case, based on the screen that appeared, I inferred that I still needed to perform another action in 'Manage Account'.''} \\ \midrule
			Improve input interaction 		& 3 & Issues related to keyboard input or interaction methods (e.g. pressing enter). & \textit{``I prefer pressing Enter instead of the application performing the action automatically.''} \\ \midrule
			UI branding to improve trust 	& 2 & Distrust caused by unfamiliar branding or third-party identity providers. & \textit{``The fact that I had to register on an unfamiliar portal made me feel somewhat distrustful.''} \\ \midrule
			Abandoned 						& 2 & The participant was unable to complete the login process or did not know how to proceed. & \textit{``I didn't know what was failing, nor where I was supposed to use the username and password.''} \\ \midrule
			Add two-factor auth. 			& 1 & Suggestions to add or improve two-factor authentication mechanisms. & \textit{``Some form of two-factor authentication or an additional method of verifying the user's identity beyond a password.''} \\ \midrule
			None 							& 47 & No suggestions provided; the login process is perceived as adequate, intuitive, or satisfactory. & \textit{``It was very easy and simple.''} \\ 
			\bottomrule
	\end{tabular}}
\end{table*}

\begin{table*}[!ht]
	\centering
	\caption{Codebook for post-task registration suggestions questionnaire.}
	\vspace{8pt}
	\resizebox{0.8\textwidth}{!}{%
		\begin{tabular}{l@{\hspace{10pt}}c@{\hspace{15pt}}p{6cm}@{\hspace{15pt}}p{6cm}}
			\toprule
			\textbf{Code} & \textbf{Freq.} & \textbf{Description} & \textbf{Example} \\
			\midrule
			Username / email confusion 		& 9 & Confusion regarding whether a username or email address should be used during registration. & \textit{``Instead of asking for a username, directly indicate that we should enter the email address.''} \\ \midrule
			More instructions / guidance 	& 9 & Requests for clearer explanations, guidance, or onboarding during registration. & \textit{``I'd like to see some simple instructions at the bottom of the main screen explaining how to register and log in to the Wi-Fi account.''} \\ \midrule
			UI visual improvement 			& 4 & Suggestions to improve visibility, layout, or readability of the registration interface. & \textit{``Indicate more clearly, using a larger font, the need to accept the policy terms and conditions.''} \\ \midrule
			Add a password alternative 		& 4 & Desire to allow password-based registration alongside passkeys. & \textit{``[...] let users set a password in case they don't want to use a passkey.''} \\ \midrule
			Add two factor auth. 			& 3 & Requests to include two-factor authentication in the registration process. & \textit{``Add double authentication.''} \\ \midrule
			Improve password policy 		& 2 & Suggestions to strengthen or clarify password requirements. & \textit{``[...] require stronger password security, such as including characters and numbers.''} \\ \midrule
			Improve security 				& 2 & General suggestions to make the registration process more secure. & \textit{``I'd like it to be more secure''} \\ \midrule
			UI branding to improve trust 	& 2 & Trust concerns related to branding, third-party services, or data handling. & \textit{``If it were affiliated with a specific institution, I'd appreciate its visual identity conveying that I'm not giving my data to third parties or subcontractors''} \\ \midrule
			Add ``see password'' button to UI 	& 1 & Suggests adding a button to see the password during registration. & \textit{``A button to view the passwords before confirming them.''} \\ \midrule
			Feedback passkey creation dialogue	& 1 & Requests feedback in the passkey creation dialogue for choosing the location. & \textit{``When registering the passkey, Android asks where it should be stored [...] I felt a bit lost at that part [...]''} \\ \midrule
			Improve speed 					& 1 & Perception that the registration process is slow or inefficient. & \textit{``[...] I'd like it to be faster [...]''} \\ \midrule
			Use biometric-based auth.		& 1 & Requests the use of biometric-based authentication. & \textit{``Fingerprint more secure [...]''} \\ \midrule
			None 							& 61 & No suggestions provided; the registration process is perceived as adequate. & \textit{``None.''} \\
			\bottomrule
	\end{tabular}}
\end{table*}

\end{document}